\documentclass[12pt,titlepage]{article}
\usepackage{amssymb,amsmath,natbib,graphicx,subcaption,changepage,relsize,amsthm,enumerate,bbm}
\usepackage[boxed]{algorithm2e}
\usepackage{tikz,url}
\usepackage[a4paper, total={6in, 9in}]{geometry}
\usepackage{setspace,titlesec}
\titlespacing{\section}{0pt}{\parskip}{0pt}
\titlespacing{\subsection}{0pt}{0pt}{0pt}
\newtheorem*{theorem}{Theorem}
\newtheorem*{proposition}{Proposition}

\newtheorem{result}{Result}
\theoremstyle{definition}
\newtheorem*{definition}{Definition}

\newcommand{\E}{\mathbb{E}}
\newcommand{\bmu}{\boldsymbol\mu}
\newcommand{\cX}{{\cal X}}
\newcommand{\bx}{\boldsymbol{x}}
\newcommand{\by}{\boldsymbol{y}}
\newcommand{\cA}{{\cal A}}
\newcommand{\cAmin}{{\cal A}^{\mathsmaller{-}}}
\newcommand{\ones}{\mathbbm{1}}
\newcommand{\vm}[1]{\mbox{\upshape vec}^\mathsmaller{-}\hspace{-0.25pc}\left(#1\right)}
\newcommand{\revvm}[1]{\mbox{\upshape rev-vec}^\mathsmaller{-}\hspace{-0.25pc}\left(#1\right)}
\newcommand{\Imin}{I^\mathsmaller{-}}
\newcommand{\Itri}{{I^\mathsmaller{\triangle}}}
\newcommand{\Prob}{\mathbb{P}}
\newcommand{\bbeta}{\boldsymbol{\beta}}
\newcommand{\bgamma}{\boldsymbol{\gamma}}
\newcommand{\btheta}{\boldsymbol{\theta}}

\newcommand{\indic}[1]{\boldsymbol{1}_{\{ #1 \}}}
\newcommand{\br}{\boldsymbol{r}}
\newcommand{\bs}{\boldsymbol{s}}
\newcommand{\mvec}{\mbox{\upshape vec}}
\newcommand{\cov}{\mbox{\upshape Cov}}
\newcommand{\eqdist}{\overset{{\cal D}}{=}}
\newcommand{\deriv}[1]{\noindent {\it Derivation:} #1 $\hfill\square$}
\newcommand{\const}{\mbox{\upshape const}}
\newcommand{\pbe}[1]{{\cal P}_{\epsilon,#1}}
\newcommand{\be}{{\cal B}_{\epsilon}}
\newcommand{\iid}{\overset{iid}{\sim}}
\newcommand{\diag}{\mbox{\upshape diag}}

\author{Daniel K. Sewell \footnote{Daniel K. Sewell is Assistant Professor, Department of Biostatistics,
University of Iowa, Iowa City, IA 52242 (E-mail: {\it daniel-sewell@uiowa.edu}).}}
\title{Simultaneous and Temporal Autoregressive Network Models}
\date{}
\begin{document}
\setlength{\abovedisplayskip}{1pt}
\setlength{\belowdisplayskip}{1pt}
\setlength{\abovedisplayshortskip}{1pt}
\setlength{\belowdisplayshortskip}{1pt}

\maketitle

\begin{abstract}
While logistic regression models are easily accessible to researchers, when applied to network data there are unrealistic assumptions made about the dependence structure of the data.  For temporal networks measured in discrete time, recent work has made good advances \citep{almquist2014logistic}, but there is still the assumption that the dyads are conditionally independent given the edge histories.  This assumption can be quite strong and is sometimes difficult to justify. If time steps are rather large, one would typically expect not only the existence of temporal dependencies among the dyads across observed time points but also the existence of simultaneous dependencies affecting how the dyads of the network co-evolve.  We propose a general observation driven model for dynamic networks which overcomes this problem by modeling both the mean and the covariance structures as functions of the edge histories using a flexible autoregressive approach.  This approach can be shown to fit into a generalized linear mixed model framework.  We propose a visualization method which provides evidence concerning the existence of simultaneous dependence. We describe a simulation study to determine the method's performance in the presence and absence of simultaneous dependence, and we analyze both a proximity network from conference attendees and a world trade network.  We also use this last data set to illustrate how simultaneous dependencies become more prominent as the time intervals become coarser.
\vspace{ 2mm}

\end{abstract}

\section{Introduction}
Co-occurrence data involves observing a set of interactions, or edges, between a set of actors.  The observed edge set and actor set together form a network object.  Such networks arise in multitudinous contexts, and the analysis of network objects has been of extreme importance to scientists in a wide range of fields.  In particular, the analysis of network dynamics is an extremely interesting and often difficult area to work in, as temporal dependencies are added to an already complex network dependence structure.

Several classes of models for temporally measured, or dynamic, networks have been proposed, mostly over the last two decades.  Each of these classes comes with pros and cons, as one would expect.  The network literature is vast even for dynamic networks, and so we only touch on a few of the key classes of models before presenting our proposed approach.  

Modeling dynamic networks using continuous-time Markov processes has a long history beginning with \cite{holland1977dynamic} and continuing with several other works \citep[e.g., ][]{wasserman1980analyzing,leenders1995models}.  A very impactful work continuing the adoption of continuous-time Markov processes is the stochastic actor-oriented model \citep{snijders1996stochastic}, which has since seen much methodological and software development \citep{rsiena}.  In this framework, each actor forms a new edge or breaks an existing edge in order to maximize that actor's so-called objective function.  This function can represent homophily on attributes or structures of the network itself, such as transitivity and reciprocity.  This class of models has been very popular and useful, and allows for wide flexibility in constructing the objective function.  

Another popular class of models used for static networks is the exponential random graph (ERG) models, proposed by \cite{frank1986markov} and developed further in countless works.  The ERG family of models was extended to dynamic networks by \cite{robins2001random}, and later extended by \cite{hanneke2010discrete} and others.  The temporal ERGM, or TERGM, in contrast to the stochastic actor-oriented model, assumes the network data to be generated according to a discrete time Markov process.  The general idea in these ERG models is to put the probabilistic structure of the observed networks in terms of functions of sufficient statistics.  These statistics often correspond to a count of some topological feature, such as triangles or $k$-stars.  The TERGM is quite flexible in the sufficient statistics that can be included in the model, is parsimonious, and can handle complex dependencies in the network.  Similar in spirit is the Separable TERGM \citep{krivitsky2014separable}, where both the formation and dissolution process are modeled.  Unfortunately, there are a variety of problems that arise with these types of ERG models.  There is the intractable normalizing constant that must be approximated, as well as degeneracy issues, or non-existence of the maximum likelihood estimators.  See, e.g., \cite{okabayashi2011parameter} and \cite{jin2013fitting} for more on this, as well as \cite{hummel2012improving} for remedies to some of these problems.  

Stochastic blockmodels \citep{holland1983stochastic,wang1987stochastic,snijders1997estimation} have been one of the most widely used and studied class of models for networks.  The mixed membership blockmodel \citep{airoldi2008mixed} was extended for dynamic networks by \cite{xing2010state}.  While quite useful, blockmodels suffer from an inability to capture network dependencies induced by complex features such as transitivity or reciprocity.

A large number of models fall into the class of latent space models.  These models originated with \cite{hoff2002latent} for static networks, and expanded in a variety of ways \citep[see, e.g.,][]{handcock2007model,krivitsky2009representing}.  These models were then extended to the dynamic context by \cite{sarkar2005dynamic}, \cite{durante2014nonparametric} and \cite{sewell2014latent}.  Scalability remains an issue with latent space models, though some attemps have been made to alleviate this \citep{raftery2012fast,salter2013variational}, and determining the dimensionality of the latent space has attracted relatively little serious work, the main exception being work done by \cite{durante2014nonparametric}.  

Our proposed work builds off of the logistic network regression models proposed by \cite{almquist2013dynamic,almquist2014logistic}.  This model provides a simple yet flexible framework for capturing the temporal dependency by modeling the mean as a function of sufficient statistics constructed from previous observations of the network.  Their model has distinct advantages such as scalability, flexibility, and easy accessibility to anyone familiar with generalized linear models.  The authors derive this model from the TERGM based on a clear set of assumptions.  The most controversial of these is that the network dyads are conditionally independent given the network history.  The problem is that the simultaneous dependence is ignored, i.e., the dependence between the co-evolving dyads.  These simultaneous dependencies play an important role in the evolution of the network, especially as the intervals at which the network is observed increase \citep{lerner2013conditional}.  It is well known that ignoring extra variation in the data can, in contexts similar to our own, lead to inconsistent estimation and attenuated estimates of the parameters \citep{demidenko2013mixed}.  Thus ignoring simultaneous dependence in the data will in many cases lead to poor estimation; we shall demonstrate this analytically in Section \ref{estimationErrors} and empirically in Section \ref{simulationStudy}.

\cite{cox1981statistical} used the terms ``parameter driven'' and ``observation driven'' models to describe two approaches for modeling binary time series data.  In the context of dynamic network analysis, we can think of the latent space approach as the analog to parameter driven models, where the temporal dependencies of the network are driven through some latent variables evolving through, say, a Markov process.  Our proposed model follows what may be considered an observation driven approach, where both the simultaneous and temporal dependencies are driven by some functions of the lagged observed networks.  More specifically, our proposed approach captures temporal dependence through modeling the mean as a function of lagged networks and similarly captures the simultaneous dependence through modeling the covariance as a function of lagged networks.  

An important motivation for this work was accessibility to appropriate network methodology for those without extensive statistical background.  We believe that those familiar with generalized linear mixed models (see Section \ref{GLMMSection}) should be able to easily understand and utilize our proposed approach, and software will be made available on the author's website to further facilitate accessibility.  While using a familiar framework, we account for both temporal and simultaneous dependence, thus avoiding the adverse inferential impacts that we otherwise would expect to occur by ignoring these two sources of variation.  

In Section \ref{methodology} we present our proposed methodology, as well as some suggestions for appropriately choosing the mean and covariance functions.  In Section \ref{estimation} we describe our approach to estimation, with the details and selected proofs given in the appendix.  Section \ref{GLMMSection} generalizes our approach by fitting our method into the familiar generalized linear mixed model framework.  In Section \ref{evidence} we describe a visualization approach to evaluating the evidence regarding the existence and impact of simultaneous dependence in the data.  In Section \ref{simulationStudy} we present a simulation study which examines the performance of our model in the presence and absence of simultaneous dependencies.  In Section \ref{dataAnalyses} we analyze two real data sets, illustrating the utility of our method and the importance of accounting for simultaneous dependence in real data, as well as illustrating how simultaneous dependence becomes more prominent as time intervals become coarser.

\section{Methodology}
\label{methodology}
\subsection{Context and notation}
We assume we have $n$ objects, or {\it actors}, each of which may have some interactions or relationships with the other actors.  If such an interaction/relationship exists between actors $i$ and $j$, we say there is an {\it edge} between them.  We assume that the set of actors are constant over time, though the edges themselves may exist during any subset of all possible time points.  Here we assume the data are collected at discrete time points.  Collectively the set of actors and the time-varying set of edges define the dynamic network.  The data obtained can then be represented by a 3-dimensional tensor, or equivalently a sequence of adjacency matrices, where each adjacency matrix, denoted as $A_t$, $t=0,1,\ldots,T$, is an $n\times n$ matrix corresponding to the edges that exist at time $t$.  That is, the $(i,j)^{th}$ entry of $A_t$, $A_{ijt}$, equals one if there is an edge from $i$ to $j$ at time $t$ and zero otherwise.  The diagonal entries of each adjacency matrix hold no meaning unless so-called self loops are allowed, that is, an actor may send an edge to itself.  
For the purposes of clarity in our exposition, we will assume in Section \ref{methodology} that such self loops are allowed as this helps facilitate the mathematical description of the model and its properties; it is trivial to translate the presented model to the context of no self loops.  However, because (1) self loops are relatively rare in practice, and (2) the derivations of our estimation algorithm requires additional non-trivial steps when self loops are not allowed, the derivations provided in our appendices assume the diagonal elements of the $A_t$'s are meaningless.  Additionally, the data in Sections \ref{simulationStudy} and \ref{dataAnalyses} do not have self loops.

We also assume there exists some exogenous covariate information with which we would like to explain or predict the edge probabilities.  These covariates may by static (e.g., race or gender) or time-varying (e.g., income or marital status).  In the remainder of the paper we will treat the covariates as though they are time-varying with the understanding that static covariates may be treated as such simply by replicating them from one time point to the next.  We denote the dyadic covariate information by the $n\times n$ matrices $X_{\ell t}$, $\ell = 1,\ldots,p_1$, $t=1,\ldots,T$.  For notational convenience, we will denote a linear combination of equal sized matrices as $\langle\bbeta,\cX_t\rangle:=\sum_{\ell=1}^{p_1}\beta_{\ell}X_{\ell t}$, where $\bbeta=(\beta_1,\ldots,\beta_{p_1})$ and $\cX_t$ is a 3-dimensional array whose $\ell^{th}$ slice is $X_{\ell t}$.  

As will be seen shortly, we shall be focusing on covariance structures, and hence it is natural to implement a probit type model for our binary dyadic data (although we will generalize the work in Section \ref{GLMMSection}).  We thus assume that there are some underlying matrices of normal random variables $A^*_{t}$ that directly correspond to $A_t$ via the surjective function $A_{ijt}=\indic{A^*_{ijt}>0}$.  

\subsection{Observation-driven model}
The proposed model is an observation-driven approach, rather than parameter-driven.  That is, we may write the conditional mean of $A^*_t$ as a function of $A_0,\ldots,A_{t-1}$ rather than as a function of some unobservable noise process.  Observation-driven approaches for temporal binary data have been well studied in simpler contexts.  While some complicated mean functions have been proposed \citep[e.g.,][]{shephard1995generalized}, often it is the simple and intuitive
$$
\E(A^*_{ijt}|A_{ij(t-1},A_{it(t-2)},\ldots) = \sum_{\ell =1}^{p_1}\beta_\ell X_{\ell t}[i,j] + \sum_{\ell=1}^{p_2}\theta_\ell A_{ij(t-\ell)},
$$
\citep[e.g.,][]{cox1981statistical,zeger1988markov} where $X[i,j]$ is the $(i,j)^{th}$ entry of the matrix $X$.  However, this simplistic mean function is insufficient for complex network objects.  With this in mind, we will allow the second term of the mean of $A_t^*$ to be $\langle\btheta,{\cal G}_t\rangle := \langle\btheta,{\cal G}(A_{t-1},A_{t-2},\ldots)\rangle$, where $\btheta=(\theta_1,\ldots,\theta_{p_2})$, and ${\cal G}_t$ maps the previous adjacency matrices onto the space of $n\times n\times p_2$ tensors, i.e., ${\cal G}_t$ uses the previous adjacency matrices to construct $p_2$ new $n\times n$ matrices.  

Note that $p_2$ does not refer to the number of lagged time points as in the simple binary time series model, but rather can encompass the number of salient features of the previous adjacency matrices, such as stability, reciprocity, or transitivity.  As a simple example, if we include stability and reciprocity for up to a lag of two time points, then $p_2=4$ and the slices of ${\cal G}_t$ are $A_{t-1}$, $A_{t-1}'$, $A_{t-2}$, and $A_{t-2}'$.  These $p_2$ covariates involving functions of the lagged network can thus be used in sophisticated ways to explain the temporal dependencies, i.e., the dependence between $A_{ijt}$ and $A_{k\ell s}$, $t\neq s$.  For examples of other ways to construct ${\cal G}_t$, see Table \ref{tableG_t} or the appendices of \cite{almquist2014logistic}.

Networks are complex objects, however, and attempting to capture all dependencies through the mean structure alone is insufficient, particularly as the intervals between time points grow larger.  One would typically expect not only the existence of {\it temporal} dependencies through which the network at varying time points are dependent, but also {\it simultaneous} dependencies which dictate how the dyads of the network co-evolve.  Thus we should be quite concerned with appropriately modeling the second moments of the $A_{ijt}^*$'s.  

With this motivation in mind, we begin with the following multivariate probit model.  Let $\cA_t$ be equal to $\mvec(A_t^*)$.  Then set 
\begin{align}
\E(A_t^*|A_{t-1},A_{t-2},\ldots) &= \langle\bbeta,\cX_t\rangle + \langle\btheta,{\cal G}_t\rangle & \label{meanAtstar}\\
\cov(\cA_t) &= \Sigma_{A^*,t}.&
\end{align}
Note that $\Sigma_{A^*,t}$ determines the covariance structure among the $n^2$ dyads, and hence has $\mathcal{O}(n^4)$ parameters.  Clearly it would not be possible to estimate such an unconstrained $\Sigma_{A^*,t}$ outside of the context of small $n$ large $T$, nor is this unconstrained covariance structure what one would expect to see in reality.  
Going to the extreme of constraining $\Sigma_{A^*,t}$ to be the identity matrix (and thus ignoring simultaneous dependence entirely) leads to the model presented in \cite{almquist2014logistic}, and hence what is presented here can be thought of as an alternative generalization of their methods (the TERGM is the original motivation for and generalization of their approach).  

\subsection{Ignoring simultaneous dependencies}
\label{estimationErrors}
Here we make a short note on estimation errors associated with ignoring existing variablity in the data.  \cite{demidenko2013mixed} gives a short discussion on these types of issues with regard to generalized linear mixed models (see chapter 7).  For our context, suppose we may write the normal random variables $A_{ijt}^*$'s as 
\[A_{ijt}^*= \langle \bbeta,\cX_t\rangle[i,j] + \langle \btheta,{\cal G}_t\rangle[i,j] + s_{it} + r_{jt} + E_{ijt},\]
where  $s_{it}$, $r_{it}$, and $E_{ijt}$ are zero mean normal random variables (possibly correlated in complex ways, though letting $s_{it},r_{it}\perp E_{ijt}\forall i,j,t$ ).  Then we have the following proposition, the proof of which is given in Appendix \ref{proofOfProposition}. 
\begin{proposition}
\begin{align}
\Prob(A_{ijt}=1|\bbeta,\btheta)&=
\boldsymbol\Phi\left(
\frac{\E(A_{ijt}^*)}{\sqrt{Var(E_{ijt})+Var(s_{it}+r_{jt})}}
\right),&
\end{align}
where $\boldsymbol\Phi(\cdot)$ is the CDF of a standard normal distribution, and $\E(A_{ijt}^*)$ is given in (\ref{meanAtstar}).  
\end{proposition}

Now consider the very simple example where we have
\[
 \begin{pmatrix}
s_{it} \\ r_{it}
\end{pmatrix} \iid N\left({\bf 0},\begin{pmatrix}
\tau_s & 0 \\ 0 & \tau_r
\end{pmatrix}\right)
\]
and constant variance for the $E_{ijt}$'s.  We can quickly see that should we ignore simultaneous dependence, any attempts to estimate $(\bbeta,\btheta)$ would in fact unintentionally lead to the attenuated estimation of $(\bbeta,\btheta)$ scaled by $Var(E_{ijt})+\tau_{s}+\tau_{r}$.  For more general cases when $Var(s_{it}+r_{jt})$ is time dependent or dependent on the actors $i$ and $j$, it is unclear what, if anything, any naive estimates of $(\bbeta,\btheta)$ are actually estimating.

\subsection{Simultaneous and temporal autoregressive model}
\label{STARModel}
A middle ground between fully ignoring simultaneous dependence and using a saturated covariance matrix $\Sigma_{A^*,t}$ would be to assume that there ought to be some connection with the covariance between two dyads and the actors that are incident on those two dyads.  This simple and intuitive idea will eventually lead us to a model resembling the social relations model \citep{warner1979new}, having the form
\[
A_{ijt}^* = \mbox{ mean structure } + \mbox{ sender effects } + \mbox{ receiver effects } + \mbox{ residuals}
\]
(the final form is given in (\ref{STAR})).  To get there, we begin by introducing the following definition.
\begin{definition}
An $n\times n$ matrix $A^*$ has a {\it role-based additive covariance structure} if 
\begin{align}\nonumber
&\cov(A^*_{ij},A^*_{k\ell}) &\\
&= \Sigma_s[i,k] + \Sigma_r[j,\ell] + \Sigma_{sr}[i,\ell] + \Sigma_{sr}[k,j]  + \sigma_R^21_{[\{(i,j)=(k,\ell)\}\cup\{(i,j)=(\ell,k)\}]} + \sigma_\epsilon^2 1_{[(i,j)=(k,\ell)]},&
\label{rbacs1}
\end{align}
where $\Sigma_s$, $\Sigma_r$, and $\Sigma_{sr}$ are $n\times n$ covariance matrices that represents respectively the covariance among the senders of the dyads, the receivers of the dyads, and between the senders and the receivers, and where $\sigma^2_R$ and $\sigma^2_\epsilon$ correspond to pair and dyad variance respectively.
\end{definition}

A role-based additive covariance structure can be interpreted to mean that the covariance between any two dyads $(i,j)$ and $(k,\ell)$ can be explained by how similar $i$ and $k$ are as senders, how similar $j$ and $\ell$ are as receivers, how $i$ and $\ell$ relate to each other as sender and receiver respectively and similarly for $k$ and $j$, the variability due to reciprocated dyads, and the inherent variability between the dyads.

The role-based additive covariance structure has a nice representation that lends itself well to estimation.  To demonstrate this, we provide the following theorem.
\begin{theorem}
The following are equivalent.  
\begin{enumerate}[(I)]
\item The $A_{ijt}^*$'s are jointly normal with a role-based additive covariance structure and mean given by (\ref{meanAtstar}).
\item $\begin{aligned}[t]
\cA_t &\sim N\Big(
\mvec(\langle\bbeta,\cX_t\rangle + \langle\btheta,{\cal G}_t\rangle),&\\
&J_n\otimes\Sigma_{st} + \Sigma_{rt}\otimes J_n + \ones_n\otimes\Sigma_{srt}\otimes\ones_n' + \ones_n'\otimes \Sigma_{srt}'\otimes\ones_n + \sigma^2_RM_R+(\sigma^2_\epsilon+\sigma^2_R)I_{n^2}
\Big),
\end{aligned}$
\begin{equation}\label{rbacs2}\end{equation}
where $\ones_k$ is the $k\times1$ vector of 1's, $J_k$ equals $\ones_k\ones_k'$, and $I_k$ is the $k\times k$ identity matrix, and where $M_R$ is a matrix such that for $1\leq i\neq j\leq n$, $M_R[ (j-1)n+i,(i-1)n+j]=1$ and $M_r[\ell,m]=0$ everywhere else.
\item $A^*_t=\langle\bbeta,\cX_t\rangle + \langle\btheta,{\cal G}_t\rangle + \bs_t\ones'+ \ones\br' + E_t,$ where
\begin{align}\nonumber 
\left(\begin{array}{c}
\bs_t\\ \br_t
\end{array}\right)&\iid N\left( {\bf 0}, 
\left(\begin{array}{cc}
\Sigma_{st} &\Sigma_{srt} \\
\Sigma_{srt}' & \Sigma_{rt}
\end{array}\right)\right),&\\ 
(E_t[i,j],E_t[j,i])'&\iid N\big({\bf 0},\sigma^2_\epsilon I_2 + \sigma^2_RJ_2\big).& \label{probitMod1}
\end{align}
\end{enumerate}
\end{theorem}
\noindent The proof is given in Appendix \ref{proofOfTheorem}.

Unconstrained, the covariance structure of (\ref{probitMod1}) still has $O(n^2)$ parameters to be estimated.  The question then is how to appropriately, yet parsimoniously, represent the covariance structure of $(\bs_t,\br_t)$.  In response, we pose the following question: if the features found in $(A_{t-1},A_{t-2},\ldots)$ can appropriately capture the temporal dependence through the mean structure, may we not also capitalize on the information stored in $(A_{t-1},A_{t-2},\ldots)$ to estimate the simultaneous dependence through the covariance structure?  \citep[This is similar in principle to ARCH models.  See][]{engle1982autoregressive}.
We propose using an autoregressive model on the covariance structure of $(\bs_t,\br_t)$ as well as on the mean structure of $A_t^*$, so that $\cov(\cA_t|\cA_{t-1},\cA_{t-2},\ldots)$ is some function of  $(\cA_{t-1},\cA_{t-2},\ldots)$.  

Specifically, we consider $\cov(\bs_t,\br_t)$ with the following structure:
 \vspace{-0.5pc} \\ 
\begin{minipage}{0.32\textwidth}
\begin{center}
\begin{equation*}
\Sigma_{st}=\sum_{k=1}^{K_s}\tau_{sk}H_{skt}
\end{equation*}
\end{center}
\end{minipage}
\begin{minipage}{0.32\textwidth}
\begin{center}
\begin{equation*}
\Sigma_{rt}=\sum_{k=1}^{K_r}\tau_{rk}H_{rkt}
\end{equation*}
\end{center}
\end{minipage}
\begin{minipage}{0.32\textwidth}
\begin{center}
\begin{equation}
\Sigma_{srt}=\sum_{k=1}^{K_{sr}}\tau_{srk}H_{srkt}
\label{covarStr1}
\end{equation}
\end{center}
\end{minipage}\vspace{0.5pc}
\\ 
\noindent
where $\tau_{sk}$, $\tau_{rk}$, and $\tau_{srk}$ are positive valued parameters, $H_{skt}$, $H_{rkt}$, and $H_{srkt}$ are functions of $(A_{t-1},A_{t-2},\ldots)$, and $H_{skt},H_{rkt}\in \mathbb{S}_+^n$ for all $k$.  Here $\mathbb{S}_+^n$ denotes the positive semi-definite (PSD) cone.  Writing $\cov(\bs_t,\br_t)$ in this manner, i.e., as a linear combination of PSD matrices, is similar in principle to covariance structures studied for many decades \citep[e.g.,][]{anderson1973asymptotically}.  Constructing the covariance matrices in this manner allows us to use the data to represent complex simultaneous dependence, while reducing the number of parameters from $O(n^2)$ to $K_s+K_r+K_{sr}$.  

Note that this does not automatically ensure that $\Sigma_{A^*,t}\in \mathbb{S}_+^{n^2}$, and so some care is still needed.  To ensure that we have a valid covariance matrix, we constrain $K_{sr} \leq \mbox{min}\{K_s,K_r\}$, and for $1\leq k \leq K_{sr}$ impose the constraint that
\begin{equation}
\left(\begin{array}{cc}
\tau_{sk}H_{skt} &\tau_{srk}H_{srkt}\\
\tau_{srk}H_{srkt}' & \tau_{rk}H_{rkt}
\end{array}\right) \in \mathbb{S}_+^{(2n)}.
\end{equation}

The structure found in (\ref{covarStr1}) allows us to further decompose $\bs_t$ and $\br_t$ as\\
\begin{minipage}{0.4\textwidth}
\begin{center}
\begin{align*}
\bs_t = \sum_{k=1}^{K_s}\bs_{kt},& \hspace{1pc} \bs_{kt} \overset{\mbox{\scriptsize ind}}{\sim} N({\bf 0},\tau_{sk}H_{skt}) \\
\br_t = \sum_{k=1}^{K_r}\br_{kt},& \hspace{1pc} \br_{kt} \overset{\mbox{\scriptsize ind}}{\sim} N({\bf 0},\tau_{rk}H_{rkt})\hspace{2pc}
\end{align*}
\end{center}
\end{minipage}
\begin{minipage}{0.55\textwidth}
\begin{center}
\begin{eqnarray}
\cov(\bs_{kt},\br_{k't})=\left\{\begin{array}{ll}
\tau_{srk}H_{srkt} &\mbox{if } 1\leq k=k'\leq K_{sr}\\
0& \mbox{otherwise.}
\end{array}\right.
\end{eqnarray}
\end{center}
\end{minipage}\vspace{0.5pc}

This then results in having our multivariate probit model with role-based additive covariance structure represented as
\begin{equation}
A_t^*= \langle\bbeta,\cX_t\rangle + \langle\btheta,{\cal G}_t\rangle +
\left(
\sum_{k=1}^{K_s} \bs_{kt}
\right)\ones' +
\ones\left(
\sum_{k=1}^{K_r}\br_{kt}
\right)' +E_t.
\label{STAR}
\end{equation}

\subsection{Broader context of sender/receiver effects}
By first assuming an intuitive form for the covariance of the dyads, we are able to arrive at a multivariate mixed effects probit model for the dynamic network, using individual sender and receiver effects.  The use of individual sender and receiver effects has a long history in network analysis, starting with \cite{warner1979new}.  In nearly all cases, the additive sender and receiver effects can be put within the framework described above by setting $K_s=K_r=K_{sr}=1$ and $H_{s1t}=H_{r1t}=H_{sr1}=I_n$.  An important work using this is the $p_2$ model of \cite{duijn2004p2}.  This work was built off of the $p_1$ model of \cite{holland1981exponential} which was not motivated by modeling an appropriate covariance structure.  Latent space models have incorporated additive sender/receiver effects as well, such as \cite{hoff2005bilinear} (which also incorporated multiplicative effects), and \cite{krivitsky2009representing}.  

The above referenced works are all concerned with static networks.  \cite{westveld2011mixed} used the ideas of sender and receiver effects to model the covariance of the data for dynamic networks.  As with the others, they constrain $K_s=K_r=K_{sr}=1$ and $H_{s1t}=H_{r1t}=H_{sr1}=I_n$, while also assuming AR processes on the sender and receiver effects (and on the residuals).  While there is merit in this approach, we still prefer capturing the temporal dependency through the observation driven model.  This is primarily because one may utilize specific network features such as stability, reciprocity, transitivity, etc. to help explain the temporal dependencies.  In this way, one may argue that there is more flexibility, and researchers can investigate the specific effects of various network features.

The way in which we use sender and receiver effects here differs in two important ways from previous uses.  First, the constraints on the covariance matrix of the dyads are relaxed to allow $\Sigma_{A^*,t}$ to be dense, thus generalizing the way that researchers have in the past used sender and receiver effects in their models.  Second, we incorporate past data to make the parameter space parsimonious.  That is, a dense covariance matrix with $O(n^4)$ unknowns can, by leveraging past information, be estimated using $K_s+K_r+K_{sr}$ parameters.  For an example of how we may do this in practice, see Section \ref{example}.

\subsection{An example of operationalization}
\label{example}
One of the strengths of (\ref{meanAtstar}) and (\ref{covarStr1}) is the flexibility in choosing the features of the previous adjacency matrices to be used in constructing the mean and covariance functions.  In this subsection we provide an example, based on sociological principles as well as previous research in statistical models for networks, with the intention that researchers using the STAR model may use whatever network features are most appropriate for their particular context.  

Fortunately for the analyst looking at dynamic network data, there has been much focus in the social science literature on the salient structures of networks.   To quote \cite{wasserman1994social},
\begin{quotation}
Many researchers have shown, using empirical studies, that social network data possess strong deviations from randomness.  $\ldots$ data often fail to agree with predictions from [models with assumptions such as equal popularity, lack of transitivity, or no reciprocity].
\end{quotation}
\cite{krackhardt2007heider} made note that it has long been argued that ``the triad, not the dyad, is the fundamental social unit that needs to be studied'' \citep[see also][]{simmel1950sociology}, which further emphasizes that transitivity is, to quote \cite{wasserman1994social} again, ``indeed a compelling force in the organization of social groups.''  

\begin{figure}
\centering
\begin{subfigure}{0.2\textwidth}
\begin{tikzpicture}[scale=1.25]
\node (i) at (0,0) {$i$};
\node (j) at (2,0) {$j$};
\node (k) at (1,1) {$k$};
\draw[->] (i) -- (k);
\draw[->] (k) -- (j);
\draw[dashed, ->] (i) -- (j);
\end{tikzpicture}
\caption{${\cal G}_{5t}[i,j]$}
\end{subfigure}
\begin{subfigure}{0.2\textwidth}
\begin{tikzpicture}[scale=1.25]
\node (i) at (0,0) {$i$};
\node (j) at (2,0) {$j$};
\node (k) at (1,1) {$k$};
\draw[->] (i) -- (k);
\draw[->] (j) -- (k);
\draw[dashed, ->] (i) -- (j);
\end{tikzpicture}
\caption{${\cal G}_{6t}[i,j]$}
\end{subfigure}
\begin{subfigure}{0.2\textwidth}
\begin{tikzpicture}[scale=1.25]
\node (i) at (0,0) {$i$};
\node (j) at (2,0) {$j$};
\node (k) at (1,1) {$k$};
\draw[->] (k) -- (i);
\draw[->] (k) -- (j);
\draw[dashed, ->] (i) -- (j);
\end{tikzpicture}
\caption{${\cal G}_{7t}[i,j]$}
\end{subfigure}
\begin{subfigure}{0.2\textwidth}
\begin{tikzpicture}[scale=1.25]
\node (i) at (0,0) {$i$};
\node (j) at (2,0) {$j$};
\node (k) at (1,1) {$k$};
\draw[->] (k) -- (i);
\draw[->] (j) -- (k);
\draw[dashed, ->] (i) -- (j);
\end{tikzpicture}
\caption{${\cal G}_{8t}[i,j]$}
\end{subfigure}
\caption{Network structures which are being summed over $k$ to determine the mean of $A_{ijt}^*$}
\label{triadFig}
\end{figure}
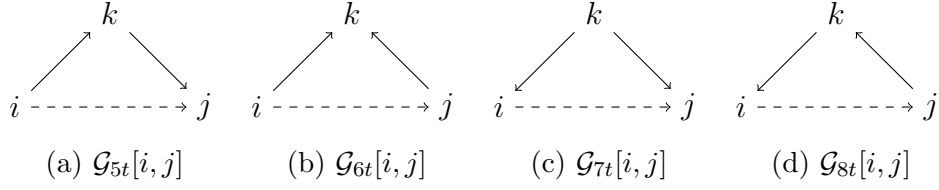

These notions then motivate the construction of ${\cal G}_t$, the 3-dimensional tensor whose $\ell^{th}$ slice is denoted by ${\cal G}_{\ell t}$, as given in Table \ref{tableG_t}.  
We can categorize these 8 structures of the network in the following terms.  ${\cal G}_{1t}$ and ${\cal G}_{2t}$ correspond to first order structures, that is, features of the network that relate to individual actors only.  ${\cal G}_{3t}$ and ${\cal G}_{4t}$ correspond to second order structures, that is, features of the network that relate to dyads.  ${\cal G}_{5t}$ to ${\cal G}_{8t}$ correspond to third order structures, that is, features of the network that relate to triads.  In particular, ${\cal G}_{5t}$ to ${\cal G}_{7t}$ correspond to transitivity in the network, i.e., the probability that a transitive relation exists, while ${\cal G}_{8t}$ corresponds to a cycle, i.e., the probability that a 3-cycle will be completed.  These last four structures are depicted visually in Figure \ref{triadFig}, where we are considering the probability of an edge from $i$ to $j$ and visualizing the transitive and cyclic triadic relations involving the third actor $k$.  One note regarding ${\cal G}_{1t}$ to ${\cal G}_{8t}$ is that these same features could of course be trivially extended to more than just a lag of 1 whenever appropriate.
\begin{table}
\begin{center}
\begin{tabular}{lll}
{\bf (out degree)}&${\cal G}_{1t}=A_{t-1}J_n$ & ${\cal G}_{1t}[i,j] = \sum_{k=1}^n A_{ik(t-1)}$ \\
{\bf (in degree)}&${\cal G}_{2t}=J_nA_{t-1}$ & ${\cal G}_{2t}[i,j] = \sum_{k=1}^n A_{kj(t-1)}$\\
{\bf (stability)}&${\cal G}_{3t}=A_{t-1}$ & ${\cal G}_{3t}[i,j] = A_{ij(t-1)}$\\
{\bf (reciprocity)}&${\cal G}_{4t}=A_{t-1}'$ & ${\cal G}_{4t}[i,j] = A_{ji(t-1)}$\\
{\bf (transitivity 1)}&${\cal G}_{5t}=A_{t-1}A_{t-1}$ & ${\cal G}_{5t}[i,j] = \sum_{k=1}^n A_{ik(t-1)}A_{kj(t-1)}$\\
{\bf (transitivity 2)}&${\cal G}_{6t}=A_{t-1}A_{t-1}'$ & ${\cal G}_{6t}[i,j] = \sum_{k=1}^n A_{ik(t-1)}A_{jk(t-1)}$\\
{\bf (transitivity 3)}&${\cal G}_{7t}=A_{t-1}'A_{t-1}$ & ${\cal G}_{7t}[i,j] = \sum_{k=1}^n A_{ki(t-1)}A_{kj(t-1)}$\\
{\bf (cycle)}&${\cal G}_{8t}=A_{t-1}'A_{t-1}'$ & ${\cal G}_{8t}[i,j] = \sum_{k=1}^n A_{ki(t-1)}A_{jk(t-1)}$
\end{tabular}
\end{center}
\caption{Example of how to construct ${\cal G}_t$, incorporating first, second and third order structures.}
\label{tableG_t}
\end{table}

Intuitively, $\Sigma_{st}$ and $\Sigma_{rt}$ ought to reflect how similar actors behave as senders and receivers respectively. We therefore suggest setting $K_s=K_r=2$, $K_{sr}=1$, and 
\begin{align}\nonumber
H_{s1t}&=H_{r1t}=H_{sr1t}=I_n&\\ 
H_{s2t}&=D_{out,(t-1)}^{-1/2}A_{t-1}A_{t-1}'D_{out,(t-1)}^{-1/2},& \label{covarStr2}\\ \nonumber
H_{r2t}&=D_{in,(t-1)}^{-1/2}A_{t-1}'A_{t-1}D_{in,(t-1)}^{-1/2},&
\end{align}
where $D_{out,(t-1)}$ and $D_{in,(t-1)}$ are diagonal matrices whose diagonal entries are the out-degrees and in-degrees of $A_{t-1}$ respectively. The $(i,j)^{th}$ entry of $H_{s2t}$ then is the number of actors to whom both $i$ and $j$ sent edges scaled by the geometric mean of the total number of actors to whom $i$ and $j$ each sent edges. In this manner we are capturing the intended notion of similarity between senders while enforcing $H_{s2t}$ to be PSD.  In fact, $H_{s2t}$ is a valid correlation matrix.  Similarly for $H_{r2t}$.  A note on the practical implementation of this is that to avoid the possibility of dividing by zero anywhere, in our analyses we set the diagonal of $A_{t-1}$ to be $\ones$ when computing $H_{s2t}$ and $H_{r2t}$.  To ensure that the covariance of $(\bs_t,\br_t)$ is PSD, and hence the covariance of $\cA_t$ is PSD, we constrain \vspace{-0.25pc}
\begin{equation}
\Omega:=
\left(\begin{array}{cc}
\tau_{s1} &\tau_{sr1} \\
\tau_{sr1} & \tau_{r1}
\end{array}\right)
 \in \mathbb{S}_+^2.
 \label{constraintS}
\end{equation}

\subsection{Undirected Networks}
\label{undirectedNetworks}
The above proposed methodology has focused on directed dynamic networks.  Simplifying to an undirected dynamic network implies that (\ref{rbacs1}) and (\ref{rbacs2}) can be written
\begin{align}\nonumber
\cov(A_{ijt}^*,A_{k\ell t}^*) &= \Sigma_{st}[i,k] + \Sigma_{st}[j,\ell]+ \Sigma_{st}[i,\ell] + \Sigma_{st}[k,j] + \sigma^21_{[(i,j)=(k,\ell)]}&\\
\Leftrightarrow \cov(\cA_t)&= J_n\otimes \Sigma_{st} + \Sigma_{st}\otimes J_n + \ones\otimes\Sigma_{st}\otimes\ones' + \ones'\otimes\Sigma_{st}\otimes\ones + \sigma^2I.
\end{align}
 The estimation algorithm given in Section \ref{estimation} can be adapted to the undirected case; some of the details which are not obvious are given in Appendix \ref{VBsolutions}.  In the analysis of Section \ref{AMD}, we set
 \begin{equation}
\Sigma_{st}=\tau_sH_{st}, \mbox{   where   } H_{st}=D_{(t-1)}^{-1/2}A_{t-1}A_{t-1}D_{(t-1)}^{-1/2}
 \end{equation}
and $D_t$ is the diagonal matrix whose diagonal entries are the degrees of the actors corresponding to $A_t$, i.e., $A_t\ones$.  For autoregressive mean terms, we used 
\begin{center}
\begin{tabular}{lll}
{\bf (degree)}&${\cal G}_{1t}=A_{t-1}J_n+J_nA_{t-1}$ & ${\cal G}_{1t}[i,j] = \sum_{k=1}^n \big(A_{ik(t-1)}+A_{jk(t-1)}\big)$ \\
{\bf (stability)}&${\cal G}_{2t}=A_{t-1}$ & ${\cal G}_{2t}[i,j] = A_{ij(t-1)}$\\
{\bf (triangle)}&${\cal G}_{3t}=A_{t-1}A_{t-1}$ & ${\cal G}_{3t}[i,j] = \sum_{k=1}^n A_{ik(t-1)}A_{jk(t-1)}$.
\end{tabular}
\end{center}

\section{Variational Bayes estimation}
\label{estimation}
From a Bayesian perspective, we would like to make posterior inference regarding the mean parameters $\bbeta$ and $\btheta$ as well as the variance components $\tau_{sk}$'s, $\tau_{rk}$'s, and $\tau_{srk}$'s.  In what follows, we will assume the particular formulation given in Section \ref{example}.  Thus of interest is deriving $\pi(\bbeta,\btheta,\Omega,\tau_{s2},\tau_{r2},\sigma^2_R|\{A_t\}_{t=0}^T)$.  Note that just as with any probit model, $\sigma^2_\epsilon$ is constrained to equal 1 for identifiability.  We assign the following priors on the model parameters.
\begin{align*}
(\bbeta',\btheta')'&\sim N({\bf 0},\mbox{diag}(\sigma^2_{\beta},\ldots,\sigma^2_{\beta},\sigma^2_{\theta},\ldots,\sigma^2_{\theta})),&\\
\tau_{s2}&\sim IG(a_{s0},b_{s0}),&\\
\tau_{r2}&\sim IG(a_{r0},b_{r0}),&\\
\Omega&\sim IW(a_{\Omega 0},B_{\Omega 0}),&\\
\sigma^2_R&\sim IG(a_{R0},b_{R0}),&
\end{align*}
where $\mbox{diag}(\sigma^2_{\beta},\ldots,\sigma^2_{\beta},\sigma^2_{\theta},\ldots,\sigma^2_{\theta})$ is the $(p_1+p_2)\times(p_1+p_2)$ diagonal matrix whose first $p_1$ diagonal entries are $\sigma^2_{\beta}$ and whose last $p_2$ diagonal entries are $\sigma^2_{\theta}$, $IG(a,b)$ is the inverse gamma distribution with shape parameter $a$ and scale parameter $b$, and $IW(a,B)$ denotes the inverse Wishart distribution with degrees of freedom $a$ and scale matrix $B$.  

Rather than implementing a computationally expensive MCMC algorithm, we implement a mean field variational Bayes (VB) algorithm.  This estimation technique finds an approximation of the posterior distribution such that the Kullback-Leibler divergence between this approximation and the true posterior distribution is minimized.  This minimization is done under the constraint that the approximated posterior density is a product of densities corresponding to a partition of the unknown model parameters.  See, e.g., \cite{gelman2004BDA3} (Chapter 13) for a brief overview of variational methods.

While much faster than MCMC, one issue with the variational Bayes algorithm is a negative bias of the variance components.  In our analyses, we found that the bias was so strong in $\sigma^2_R$ as to render the reciprocity effects negligible, which led to poorer performance overall.  To address this, first consider further data augmentation via the $n\times n$ symmetric matrices of dyad-pair specific random effects $R_t$, such that $R_t[i,j]=R_t[j,i]\iid N(0,\sigma^2_R)$.  That is, we now have the equivalent form of (\ref{STAR})
\begin{equation}
A_t^*= \langle\bbeta,\cX_t\rangle + \langle\btheta,{\cal G}_t\rangle +
\left(
\sum_{k=1}^{K_s} \bs_{kt}
\right)\ones' +
\ones\left(
\sum_{k=1}^{K_r}\br_{kt}
\right)' +R_t + \widetilde E_t,
\label{STARWthiRecRE}
\end{equation}
where $\widetilde E_t$ is a matrix of $iid$ normal random variables with zero mean and variance $\sigma^2_\epsilon$.  To prohibit $\sigma^2_R$ from shrinking to zero, we treat it as a hyperparameter for the $R_t$'s.  While not ideal, this seemed to improve overall performance.

The specific form of the approximated posterior is
\begin{align}\nonumber
&\pi(\bbeta,\btheta,\tau_{s2},\tau_{r2},\Omega,\sigma^2_R,\{A_t^*\}_{t=1}^T,\{\bs_{1t},\br_{1t},\bs_{2t},\br_{2t}\}_{t=1}^T,\{R_t\}_{t=1}^T|\{A_t\}_{t=0}^T) &\\
\approx & \hspace{0.1pc}q_1(\bbeta,\btheta)q_2(\tau_{s2},\tau_{r2},\Omega)q_3(\{\cA_t\}_{t=1}^T)q_4(\{\bs_{1t},\br_{1t},\bs_{2t},\br_{2t}\}_{t=1}^T)q_5(\{R_t\}_{t=1}^T)q_6(\sigma^2_R).&
\end{align}
This is an iterative scheme, in which we use the parameters from, say, $q_{\ell}$ to estimate $q_{m}$ and vice versa.  The closed-form solutions to the VB updates are given in Appendix \ref{VBsolutions}.  The derivations for the sender and receiver effects are also provided, as these are not straightforward due to the fact that the derivations must be taken with respect to the distribution of $A_t^*\circ (J_n-I_n)$ rather than $A_t^*$, as given in (\ref{STAR}).

The variational Bayes approach is quite fast and yields good point estimates.  This comes at a cost, however.  Variational Bayes algorithms may get stuck in local modes, and which local mode one ends up in may be highly dependent on the starting values \citep[see, e.g.,][for more detailed studies using variational approaches]{bickel2013asymptotic,salter2013variational}.  Additionally, by partitioning the parameters and forcing them to be independent in the approximate posterior, the posterior probability regions are typically much too concentrated.  In our context we found that a Gibbs sampler obtained similar posterior means, though wider credible intervals.  The MCMC algorithm was simply too slow in practice for networks of medium to large size, however.

\section{Generalizing to weighted networks}
\label{GLMMSection}
In this section we demonstrate how to generalize our approach to weighted networks in which the dyads are not constrained to $\{0,1\}$.  We accomplish this by placing our work within the framework of a generalized linear mixed model (GLMM). Most researchers, statisticians or not, are familiar with GLMMs which are often the tool of choice for modeling dependent non-Gaussian data.  The general framework assumes that a function of the means of the random variables are themselves correlated (typically Gaussian) random variables, thus allowing researchers to control for the correlation among the data.  Specifically, for some response vector $\by$, covariate matrix $X$, random variables $\bgamma$, and design matrix $Z$ we write 
\begin{equation}
g\Big(\E(\by)\Big) = X\bbeta + Z\bgamma.
\label{GLMMGeneral}
\end{equation}
(Note that the notation in (\ref{GLMMGeneral}) is not linked to anything previously given, but is rather a general form for a GLMM).  

Up to this point we have assumed a probit model, as this was a natural approach to dealing with complex dependencies in binary data.  This is equivalent to a GLMM using the normal inverse cumulative distribution function as the link function $g$.  Placing our proposed methods within the GLMM framework allows us to use other link functions such as a logit() for logistic regression, as well as allowing us to model other types of non-Gaussian data; e.g., should our network data be count, as is often the case, we may use a log link corresponding to a Poisson or Negative Binomial family of distributions.  Countless texts describe these models, and in fact GLMMs are so prevalent that many fields have books or articles demonstrating how to apply GLMMs to their specific subject area \cite[e.g.,][]{bolker2009generalized,gbur2012analysis,krueger2014modeling,bharadwaj2016generalized}.

We wish to maintain the covariance structures detailed in Section \ref{STARModel}, and in particular that implied by (\ref{STARWthiRecRE}) but generalize it to other link functions and other data types.  This can be done by setting
\begin{align}\nonumber
&g\Big(\E({\cal A}_t|{\cal A}_{t-1},{\cal A}_{t-2},\ldots)\Big) &\\ \nonumber
&=  (\vm{X_{1t}},\vm{X_{2t}},\ldots,\vm{{\cal G}_{1t}},\vm{{\cal G}_{2t}},\ldots)\footnotesize\begin{pmatrix}
\bbeta \\ \btheta
\end{pmatrix}\normalsize + Z\bgamma_t, \\ \nonumber
Z&=\begin{pmatrix}
\ones_{\mathsmaller{K_s}}'\otimes Z_{s} & \ones_{\mathsmaller{K_r}}'\otimes Z_{r}&Z_{rec}
\end{pmatrix}, \\
\bgamma_t&=\begin{pmatrix}
\bs_{1t}' & \cdots & \bs_{\mathsmaller{K_s}t}' & \br_{1t}' & \cdots & \br_{\mathsmaller{K_r}t}' & {\cal R}_t' &\\
\end{pmatrix}',
\end{align}
where ${\cal R}_t$ contains the lower triangular elements of $R_t$ (i.e., ${\cal R}_t = (R_{21t},R_{31t},\ldots,R_{n(n-1)t})$), and where $\vm{M}$ for some $n\times n$ square matrix $M$ is the standard $\mvec(M)$ while omitting the diagonals; hence $\vm{M}$ will be an $n(n-1)\times1$ vector.  To construct $Z_s$, we may stack $I_{n,(-1,\cdot)}$, $I_{n,(-2,\cdot)}$, $\cdots$, and $I_{n,(-n,\cdot)}$ to form a $n(n-1)\times n$ matrix, where $I_{n,(-i,\cdot)}$ is the $n\times n$ identity matrix with the $i^{th}$ row removed.  $Z_r$ is simply $I_{n}\otimes \ones_{n-1}$.  Constructing the $n(n-1)\times n(n-1)/2$ matrix $Z_{rec}$ is perhaps the most involved, but can be accomplished by the following pseudocode:

\begin{algorithm}[H]
Set all elements of $Z_{rec}$ to 0.
\BlankLine
\For{$i \in \{1,2,\ldots,n\}$}{
	\For{$j \in \{1,2,\ldots,n\}\setminus i$}{
		$r \leftarrow (n-1)(j-1) + i - 1_{[i>j]}$
		
		\lIf{$i>j$}{$c=n(j-1)- \frac{j(j+1)}{2}+i$}
		\lElse{$c=n(i-1)- \frac{i(i+1)}{2}+j$}
		
		$Z_{rec}[r,c]\leftarrow 1$
	}
}
\end{algorithm}

\vspace{0.5pc}
By placing our methods within the GLMM framework we provide an easy way to handle a wide range of data types as well as overdispersion.

\section{Evidence of simultaneous dependence}
\label{evidence}
We now begin to address determining whether or not simultaneous dependence exists.  Just as with mixed models, we could check the intraclass correlation between the pairs of residuals $E_t[i,j]$ and $E_t[j,i]$ to evaluate the importance of simultaneous reciprocity.  That is, estimate 
\begin{equation}
\frac{\sigma_R^2}{\sigma_R^2+1}.
\label{ICCForReciprocity}
\end{equation}
The issue is not so straighforward for the other types of simultaneous dependence.  Consider the case where the variance of $A_{ijt}^*$ does not depend on the actors $i$ and $j$ nor the time $t$, the off diagonals of $H_{srk}$ are 0 for all $k$, and the $H_{sk}$'s and $H_{rk}$'s have been scaled such that the diagonal entries are 1 (as is true in our example of Section \ref{example}).  Then analogously to (\ref{ICCForReciprocity}), one may consider the vector
\begin{align}
&{\bf v}/({\bf v}'\ones) \hspace{1pc}\mbox{where}\hspace{1pc}
{\bf v}=(\tau_{s1},\tau_{s2},\ldots,\tau_{sK_s},\tau_{r1},\ldots,\tau_{rK_r},\sigma^2_R,1).&
\label{notICC}
\end{align}
Though (\ref{notICC}) appears similar to a vector of intraclass correlations, these two things are in fact not comparable.  (\ref{notICC}) is only a ratio of variance components, while (\ref{ICCForReciprocity}) is a veritable correlation.  In the context of a directed network, there are seven correlations we could consider: $Cor(A_{ijt}^*,A_{k\ell t}^*)$, $Cor(A_{ijt}^*,A_{ki t}^*)$, $Cor(A_{ijt}^*,A_{kj t}^*)$, $Cor(A_{ijt}^*,A_{i\ell t}^*)$, $Cor(A_{ijt}^*,A_{ijt}^*)$,$Cor(A_{ijt}^*,A_{j\ell t}^*)$, and $Cor(A_{ijt}^*,A_{jit}^*)$.  Moreover, these seven correlations very well may differ based on which actors we are considering!  Instead, we present a visualization method that may be used to assess the evidence regarding the existence and impact of simultaneous dependence.  

The main idea is that we would like to evaluate how much of our posterior distributions of $\left(\{\bs_{kt}\}_{k=1}^{K_s},\{\br_{kt}\}_{k=1}^{K_r} \right)$, $t=1,\ldots,T$, are located within some small ball around zero.  If there is no simultaneous dependence, then we would expect the posterior distributions to reflect this in having most of their mass near zero.  Hence we are concerned with
\begin{align}\nonumber
\pbe{t} &:= \int_{\be} dF\left(\{\bs_{kt}\}_{k=1}^{K_s},\{\br_{kt}\}_{k=1}^{K_r} \hspace{0.25pc}  |  \hspace{0.25pc} \{A_t\}_{t=1}^T\right) &\\
&= \Prob\Big(\big\|(\bs_{1t}',\ldots,\bs_{K_st}',\br_{1t}',\ldots,\br_{K_rt}') \big\| < \epsilon \hspace{0.25pc}  | \hspace{0.25pc}  \{A_t\}_{t=1}^T\Big),&
\label{postBall1}
\end{align}
where $\be$ represents the ball around zero of radius $\epsilon$.  This probability is very easily and accurately estimated using a Monte Carlo approximation using draws from $q_4$.  We can then plot $\pbe{t}$ vs. $\epsilon$ to obtain a visualization of the magnitude of our individual effects at each time point.

Our estimate of this high dimensional posterior distribution, $q_4$, has the surprising characteristic that most of the probability mass lies within a thin shell far from the posterior mean (intuitively, this is because the volume of $\be$ grows exponentially with $n$).  Therefore we need some comparison for the $\pbe{t}$'s.  It may be helpful to compare the posterior for $\big\|(\bs_{1t}',\ldots,\bs_{K_st}',\br_{1t}',\ldots,\br_{K_rt}') \big\| $ with the distribution of the magnitude of a $N({\bf 0},\frac{p(\sigma^2_R+1)}{(1-p)(K_s+K_r)}I_{n(K_s+K_r)})$ random variable for some $p\in(0,1)$.  The distribution of this comparative random variable arises from letting the ratio of variances in (\ref{notICC}) sum to a proportion $p$ for these simultaneous dependence terms (and letting each of the $K_s+K_r$ terms contribute equally);  that is, what does the distribution of $\big\|(\bs_{1t}',\ldots,\bs_{K_st}',\br_{1t}',\ldots,\br_{K_rt}') \big\|$ look like if simultaneous dependence accounts for $p(100)\%$ of the variance of the $A_{ijt}^*$'s compared with the inherent noise?  Though there well may be better comparative distributions, what we have described provides a reasonable frame of reference by which we may evaluate the strength of the evidence of simultaneous dependence as given by the posterior distribution for the sender and receiver effects.  By looking at the visualization rather than just the ratio of variance components, we do not throw away the effects of the off-diagonal elements of the covariance matrices $\Sigma_{st}$ and $\Sigma_{rt}$ nor the entirety of $\Sigma_{srt}$ when evaluating the evidence of the existence of simultaneous dependence.

The distribution of the magnitude of the comparative random variable can be evaluated in the following way.  Let $\bx\sim N_n({\bf 0},\sigma^2 I_n)$ (e.g., $\sigma^2 =p(\sigma^2_R+ 1)/((1-p)(K_s+K_r))$).  Then let $Y^2:=\bx'\bx/\sigma^2\sim \chi^2(n)$.  Then $Y\sim \chi(n)$ and thus 
\begin{equation}
\Prob(\|\bx\|\leq \epsilon)= 
\Prob(Y\leq \frac{\epsilon}{\sigma}) = \frac{\gamma(n/2,(\epsilon/\sigma)^2/2)}{\Gamma(n/2)},
\label{chiDist}
\end{equation}
where $\gamma(\cdot,\cdot)$ is the lower incomplete gamma function.  Using this we can directly compute ${\cal P}_\epsilon$ corresponding to this comparative random variable.  

\begin{figure}
\centering
\begin{subfigure}{0.48\textwidth}
\includegraphics[width=\textwidth]{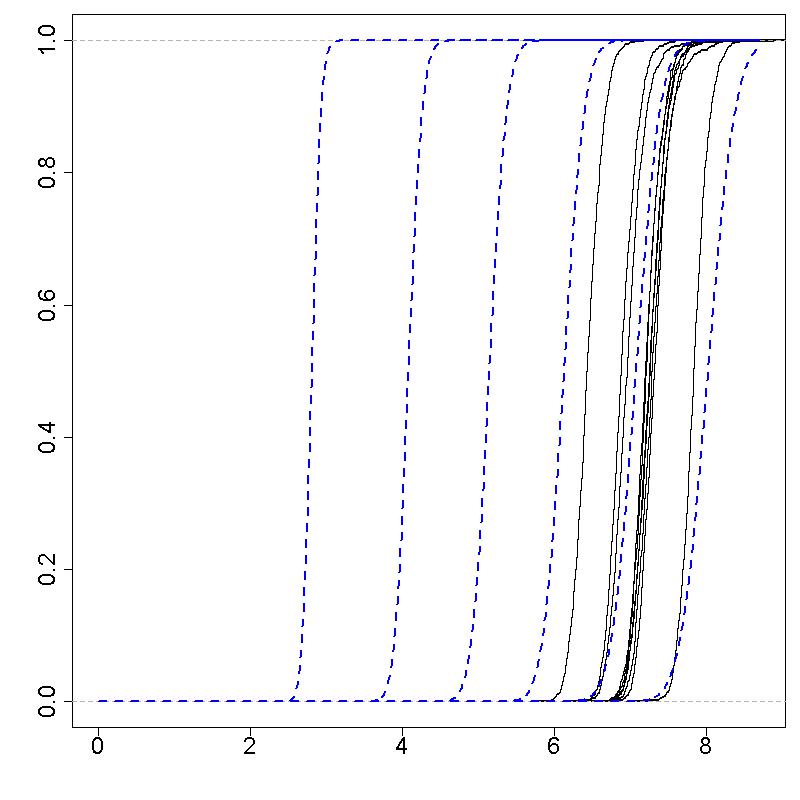}
\end{subfigure}
\begin{subfigure}{0.48\textwidth}
\includegraphics[width=\textwidth]{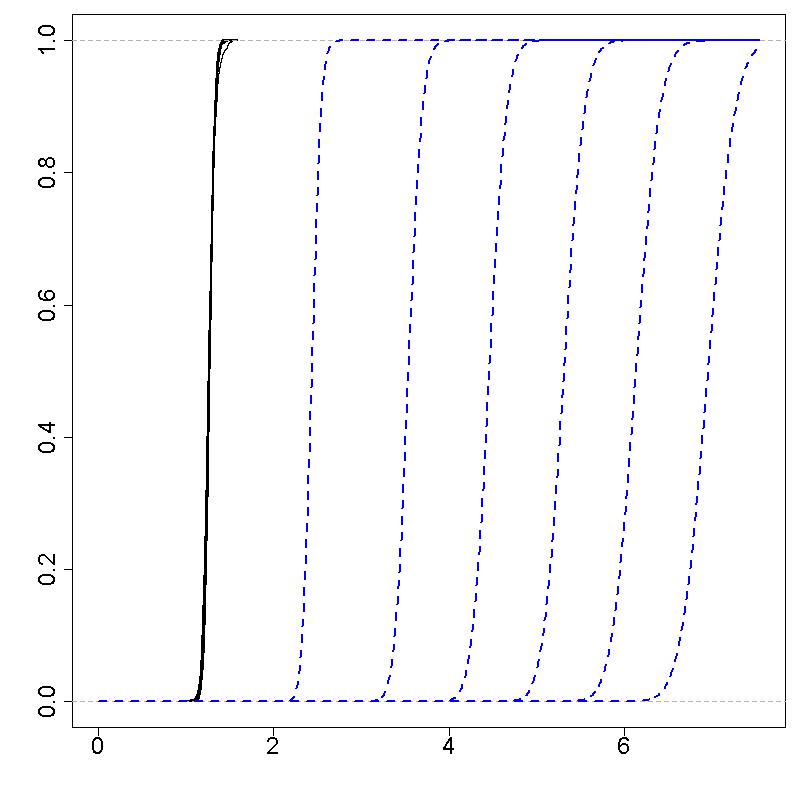}
\end{subfigure}
\caption{Empirical example of the visualization of the existence of simultaneous dependence.  The horizontal axis corresponds to the $\epsilon$ radius of a ball $\be$ about zero, and the vertical axis is $\pbe{\cdot}$.  Each solid line corresponds to a time point ($T=10$), and the dotted lines correspond to the comparative random variable having  proportion of variance attributable to simultaneous dependence of, from left to right, $p=0.05,0.1,0.15,0.2,0.25,0.3$.  The left panel corresponds to data generated with simultaneous dependence and the right panel without.}
\label{RETestEx}
\end{figure}

Figure \ref{RETestEx} provides an empirical demonstration of the proposed visualization technique using the results from an arbitrarily chosen simulated data set as described in Section \ref{simulationStudy}; note that we used the variance of the estimated $R_t$'s as a proxy for $\sigma^2_R$.  The left panel corresponds to data generated with simultaneous dependence and the right panel without.  The solid lines correspond to the individual effects at a particular time point, and the dotted lines correspond to the comparative noise for $p\in\{0.05,0.1,0.15,0.2,0.25,0.3\}$.

\section{Simulation study}
\label{simulationStudy}
We performed a simulation study in order to investigate two things.  First, what is the effect of ignoring simultaneous dependence when it exists?  Second, what is the effect of modeling simultaneous dependence when it does not exist?  Specifically, we wish to investigate the effects on the mean parameters, as these will typically be the parameters of interest to the researcher.  To this end, we simulated 100 network data sets where there was simultaneous dependence and 100 without such dependencies.  For each of these 200 data sets we fit two models, one accounting for and the other ignoring these dependencies.

Each simulated data set had $n=100$ and $T=10$.  We incorporated two covariates as well as an intercept (i.e., $p_1=3$).  The first dyadic covariate was a binary variable taking values 0 or 1 with equal probability; this covariate was treated as constant over time.  The second covariate was constructed by first simulating $n$ AR(1) processes with autoregressive coefficient equal to 0.9 and transition variance equal to 0.05, and then at each time point taking the distance between the corresponding cross-sectional views of the AR(1) time series.  The coefficients were then set to be $\bbeta=(-2.5,0.5,-2)$ for the intercept, first covariate, and second covariate respectively.  We set $\btheta=(0.0075,0.0075,0.75,0.75,0.025,0.025,0.025,-0.05)$, corresponding to ${\cal G}_{1t}, \ldots, {\cal G}_{8t}$ respectively, where the ${\cal G}_{\ell t}$'s are as given in Section \ref{example}.  Note that $\btheta_3$ and $\btheta_4$ needed to be on different scales, as these were the only coefficients corresponding to network structures taking values in $\{0,1\}$ rather than $\{0,1,\ldots,n-1\}$.  For the simulations with simultaneous dependence we set $\tau_{s2}=0.2$, $\tau_{r2}=0.1$, the diagonal of $\Omega$ to be $(0.25,0.5)$, the off-diagonals of $\Omega$ equal to $0.1$, and $\sigma^2_R=0.5$.

\begin{figure}[t]
\centering
\begin{subfigure}{0.45\textwidth}
\includegraphics[width=\textwidth]{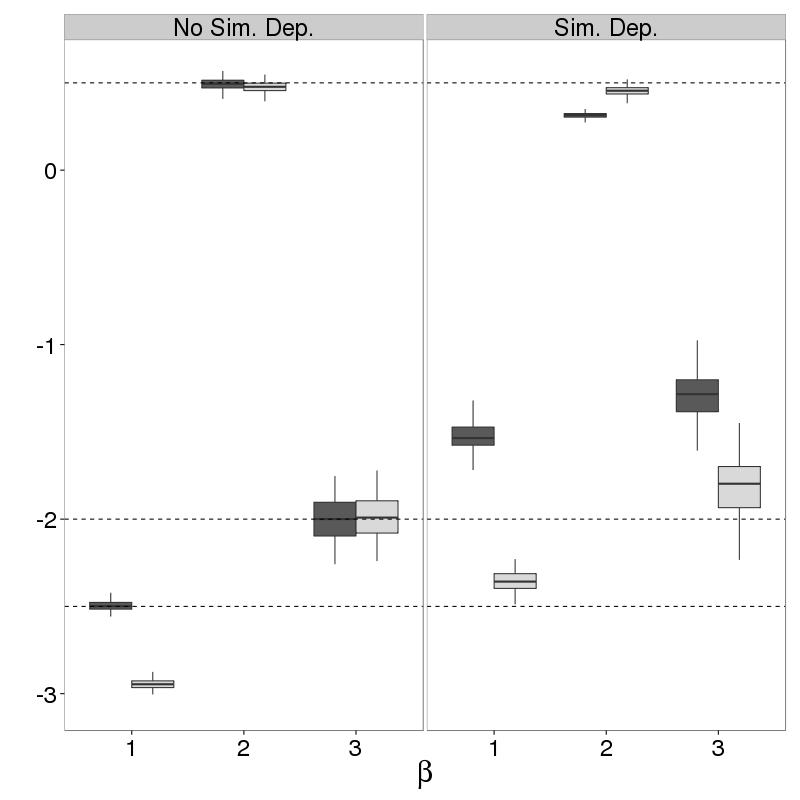}
\caption{}
\label{simBeta}
\end{subfigure}\hspace{2pc}
\begin{subfigure}{0.45\textwidth}
\includegraphics[width=\textwidth]{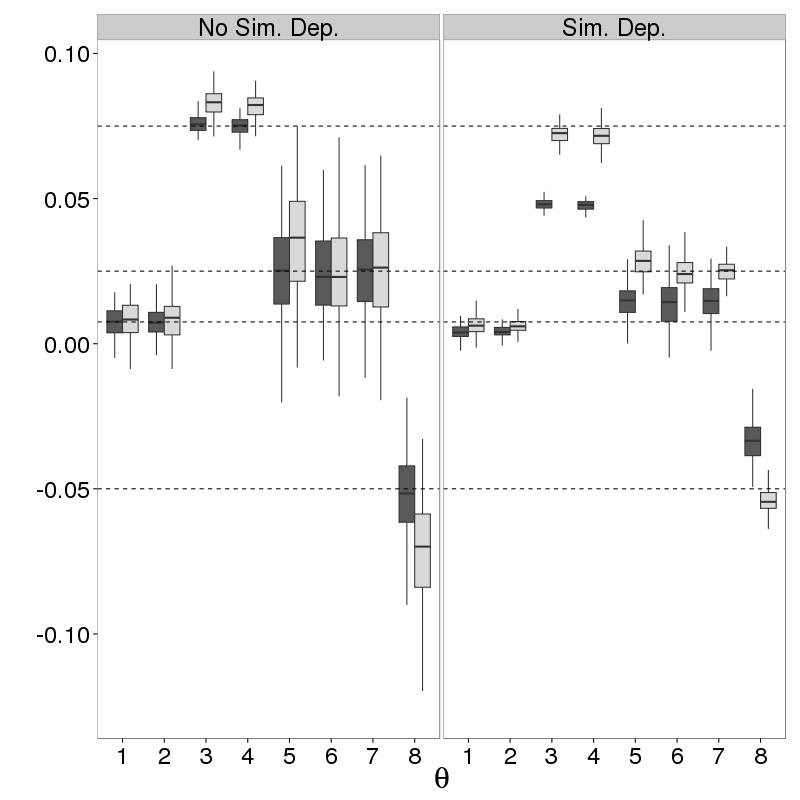}
\caption{}
\label{simTheta}
\end{subfigure}
\caption{Posterior means of (a) $\bbeta$ and (b) $\btheta$ from analyzing the simulated datasets described in Section \ref{simulationStudy}.  Note that $\btheta_3$ and $\btheta_4$ have been scaled by 1/10 for visualization purposes.  Horizontal dotted lines indicate true values of the parameters; the true $\bbeta$ equals $(-2.5,0.5,-2)$, and the true $\btheta$ equals $(0.0075,0.0075,0.75,0.75,0.025,0.025,0.025,-0.05)$.  Lightly shaded boxplots correspond to accounting for simultaneous dependence in the model; dark shaded boxplots correspond to ignoring the simultaneous dependence.}
\label{simBetaTheta}
\end{figure}

The results are given graphically in Figure \ref{simBetaTheta}.  Figure \ref{simBeta} shows the boxplots of the estimates of the $3\times1$ vector $\bbeta$.  The columns correspond to the true model, and the shade of the boxplots correspond to whether or not simultaneous dependence was accounted for.  From this we see that in the presence of simultaneous dependence, our proposed approach does a much better job at estimating the true values of $\bbeta$ than when the simultaneous dependence is ignored.  In the absence of simultaneous dependence, with the exception of the intercept (arguably of little importance in most research settings) our proposed approach performs very comparably to the models which ignore simultaneous dependence.  We can reach the same conclusions looking at Figure \ref{simTheta}, which gives the boxplots of the estimates of the $8\times1$ vector $\btheta$.  

In summary, accounting for simultaneous dependence in the model is extremely important in obtaining more accurate estimates of the coefficients in the mean function, and doing so even in the absence of simultaneous dependence does not seem to do much harm in the estimation.  If concerns persist, one may perform the visualization described previously, as seen in Figure \ref{RETestEx}, to determine whether or not to include simultaneous dependence in the final model.  

\section{Data analyses}
\label{dataAnalyses}
We now look at two real data sets with the intent of illustrating how our approach can be implemented in practice both for directed and undirected data.  In the last example we illustrate the change in impact from simultaneous dependence as the time intervals vary from fine to coarse. 

\subsection{Conference proximity network}
\label{AMD}
We first look at a proximity network taken from conference goers at The Last Hope Conference, collected and made available by the OpenAMD Project \citep{amdhope}.  The 2008 conference goers had the option to wear an RFID badge which tracked their movements throughout the conference.  Thus we are able to construct a proximity network, connecting two actors if they spent time close to one another.  This type of network is quite important in, e.g., infectious disease \citep{vanhems2013estimating} and the study of human behavior and organization \citep{eagle2006reality}.  Our undirected network data consisted of 1,190 actors over 29 hours (i.e., $T=29$).  We set $A_{ijt} (=A_{jit})$ to be 1 if actors $i$ and $j$ visited the same location during the $t^{th}$ hour.  

\begin{figure}[t]
\centering
\begin{subfigure}{0.45\textwidth}
\includegraphics[width=\textwidth]{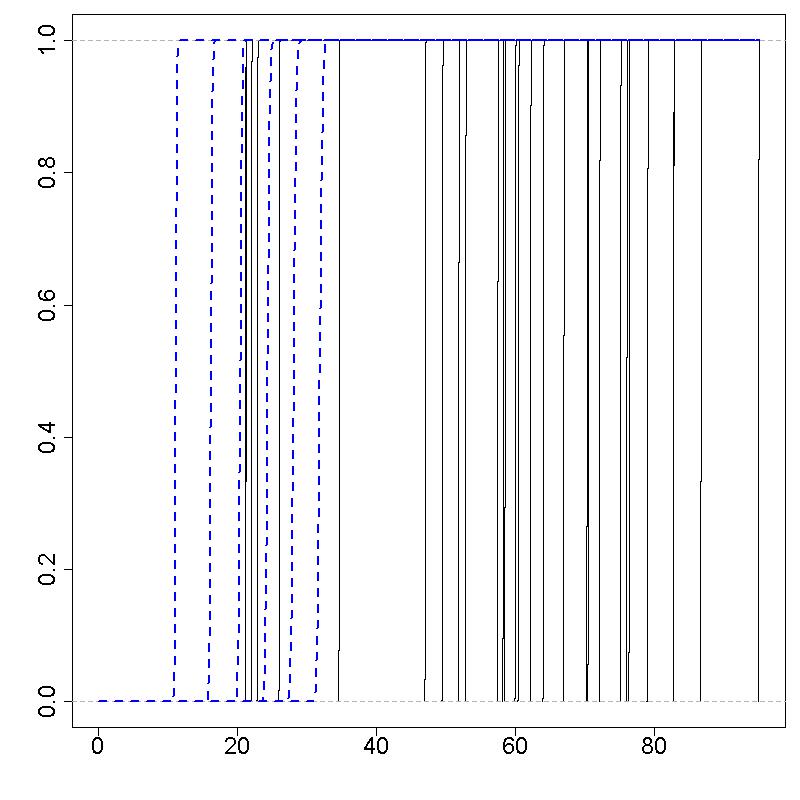}
\caption{Plot of $\pbe{t}$ vs. $\epsilon$.  Each solid curve corresponds to the individual effects from a particular time point.  The dotted lines correspond to the comparative random variable setting $p=0.05,0.1,0.15,0.2,0.25,0.3$.  See Section \ref{evidence} for details.}
\label{AMD_REEval}
\end{subfigure} \hspace{2pc}
\begin{subfigure}{0.45\textwidth}
\includegraphics[width=\textwidth]{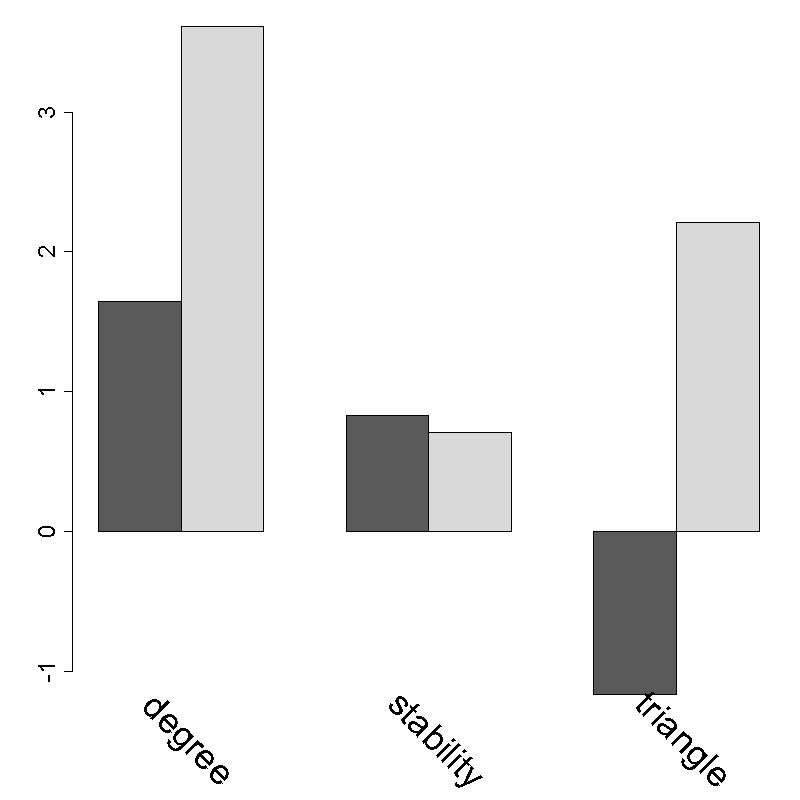}
\caption{Posterior means for the coefficients of $\btheta$.  Dark gray indicates ignoring simultaneous dependence, while light gray indicates accounting for this dependence in the model.}
\label{AMD_Thetas}
\end{subfigure}
\caption{Results from the AMD proximity network data}
\end{figure}

Figure \ref{AMD_REEval} shows the evidence of simultaneous dependence.  From this plot we see that there is very strong evidence of such dependencies even though the time intervals are rather fine (1 hour).  Figure \ref{AMD_Thetas} shows the posterior means for the autoregressive terms when ignoring simultaneous dependence (dark gray) and when accounting for it (light gray).  Notice that the estimates are, with the exception of stability, quite different; indeed, ignoring simultaneous dependence leads to a negative estimate for the effect of triangles, which seems very unlikely given previous work done on structural balance theory.

\subsection{World trade data}
\label{worldTrade}
The second data set that we consider here is that of a world trade network.  We let $A_{ijt}$ be 1 if country $i$ exports to country $j$ at time $t$.  This data were collected from the Correlates of War Project \citep{barbieri2012correlates,barbieri2009trading}.  Along with the export/import data, we used as covariates religious makeup of a country \citep{maoz2013world}, defense pacts, neutrality pacts, non-aggression pacts, and ententes \citep{gibler2009international}.  We analyze this data in two ways.  First, we focus on a larger number of countries that exist over recent years.  We then look at a smaller subset of countries that all exist over a longer period of time and look at how the evidence for simultaneous dependence changes as the time intervals get coarser.

\subsubsection{179 nations from 1993 to 2009}
We consider all countries that exist and are involved in trade on an annual basis over the period from 1993 to 2009.  For each of these countries we have the measurements of the proportion of their population that belongs to each of the main world religions and the sub-branches of these religions (a total of 30 categories).  These measurements only occur once every 5 years which we interpolated to construct annual religious data.  We then constructed the dyadic covariates by taking the Hellinger distance of two multinomial distributions whose probability vectors equal those nations' vector of proportions of religious adherents.  Letting $p_{it}$ be the $30\times1$ vector of the $i^{th}$ nation's proportion of religious adherents, this is equivalent to setting the dyadic covariate between $i$ and $j$ equal to $\sqrt{1-\sum_{r=1}^{30}\sqrt{p_{itr}p_{jtr}}}$.  The four types of pacts each were simply binary variables indicating whether or not countries $i$ and $j$ were engaged in such a pact during year $t$.

\begin{figure}[p]
\centering
\begin{subfigure}{0.45\textwidth}
\includegraphics[width=\textwidth]{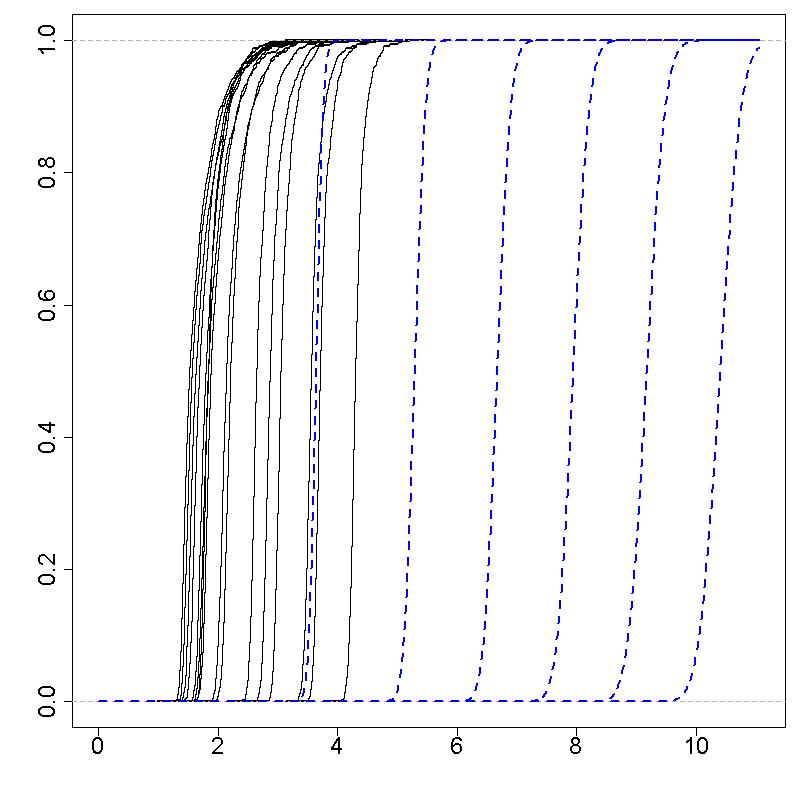}
\caption{Plot of $\pbe{t}$ vs. $\epsilon$.  Each curve corresponds to the random effects from a particular time point.  The dotted lines correspond to the comparative random variable setting $p=0.05,0.1,0.15,0.2,0.25,0.3$.  See Section \ref{evidence} for details.}
\label{COWRETest}
\end{subfigure}\hspace{2pc}
\begin{subfigure}{0.45\textwidth}
\includegraphics[width=\textwidth]{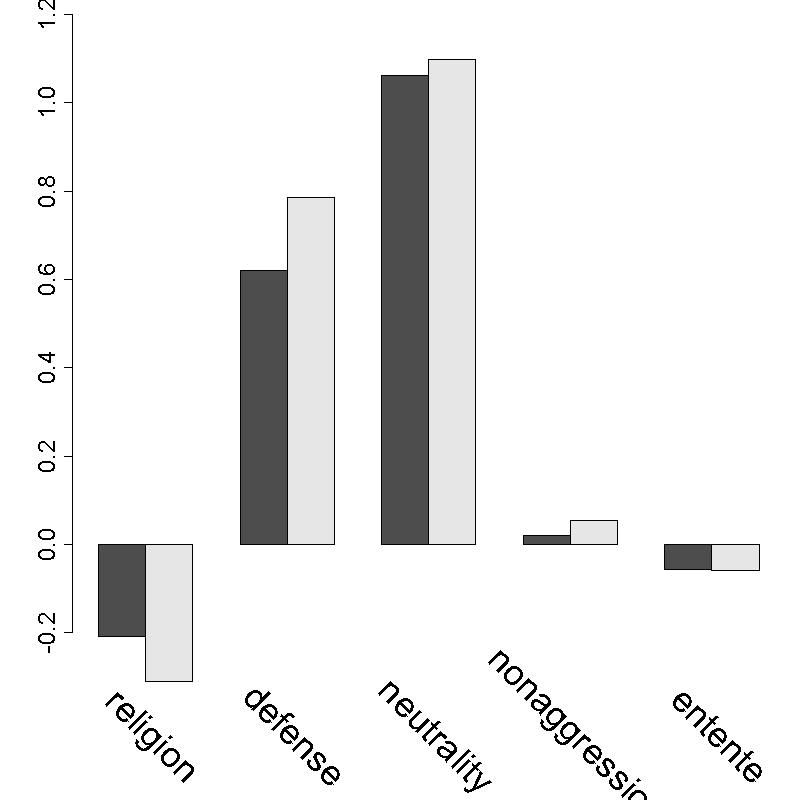}
\caption{Posterior means of the covariates ($\bbeta$).  Dark gray indicates ignoring simultaneous dependence, while light gray indicates accounting for this dependence in the model.}
\label{COWEffectSizeBeta}
\end{subfigure}\\
\begin{subfigure}{0.45\textwidth}
\includegraphics[width=\textwidth]{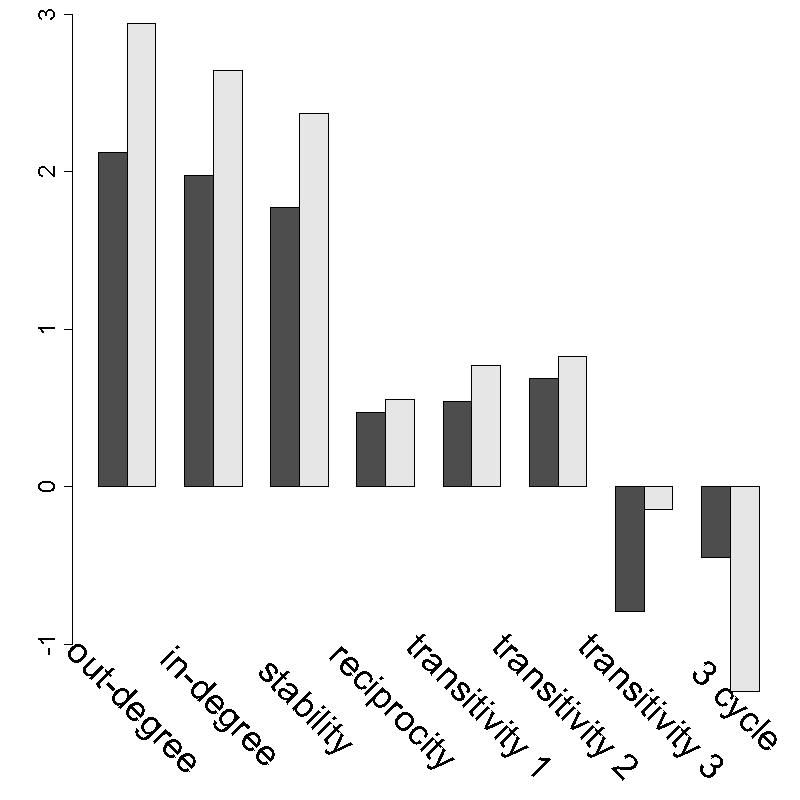}
\caption{Posterior means of the autoregressive terms ($\btheta$).  Dark gray indicates ignoring simultaneous dependence, while light gray indicates accounting for this dependence in the model.}
\label{COWEffectSizeTheta}
\end{subfigure}

\caption{Results from the world trade network data}
\end{figure}

Figure \ref{COWRETest} depicts the evidence of simultaneous dependence.  From this we see that we there is evidence of non-negligible simultaneous dependence, though much less so than in the AMD network data.  Figure \ref{COWEffectSizeBeta} shows the posterior means for the covariates and Figure \ref{COWEffectSizeTheta} shows the same for the autoregressive terms, where again dark gray indicates ignoring simultaneous dependence and light gray indicates accounting for it in the model.  As is consistent with the simulation results, when there is weaker simultaneous dependence in the data, these estimates are more in agreement.  There are still some differences, mostly manifested in the attenuation of the estimates as well as more dramatic differences in the triadic effects.

\subsubsection{Evaluating the effect of the time interval on simultaneous dependence}
As we have just seen, even at annual increments we see the presence of simultaneous dependence.  We now show how this presence increases as the time intervals become coarser.  We now consider the time interval from 1900 to 2000.  This naturally diminishes the number of nations that exist during the entirety of the specified time interval, and we are left with 28 nations.  We apply our model to these 28 nations looking at every year, every 5 years, every 10 years, every 20 years, and every 25 years.  Intuition \citep[as well as previous work by][]{lerner2013conditional} tells us that the simultaneous dependence should grow as the time interval becomes larger, and in fact this is what we see.  

\begin{figure}[t]
\centering
\begin{subfigure}{0.3\textwidth}
\includegraphics[width=\textwidth]{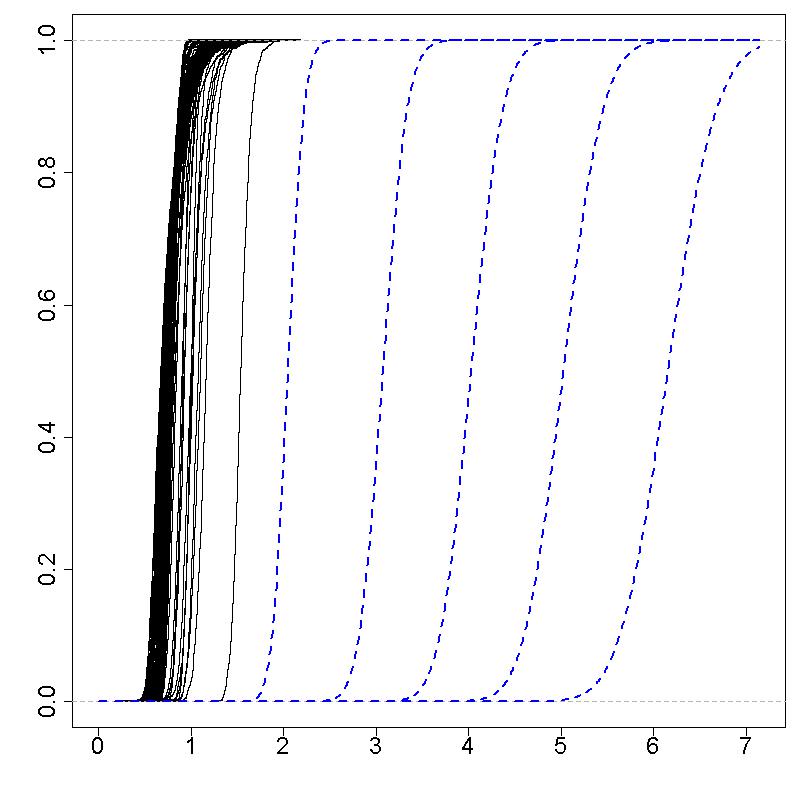}
\caption{Annual}
\end{subfigure}
\begin{subfigure}{0.3\textwidth}
\includegraphics[width=\textwidth]{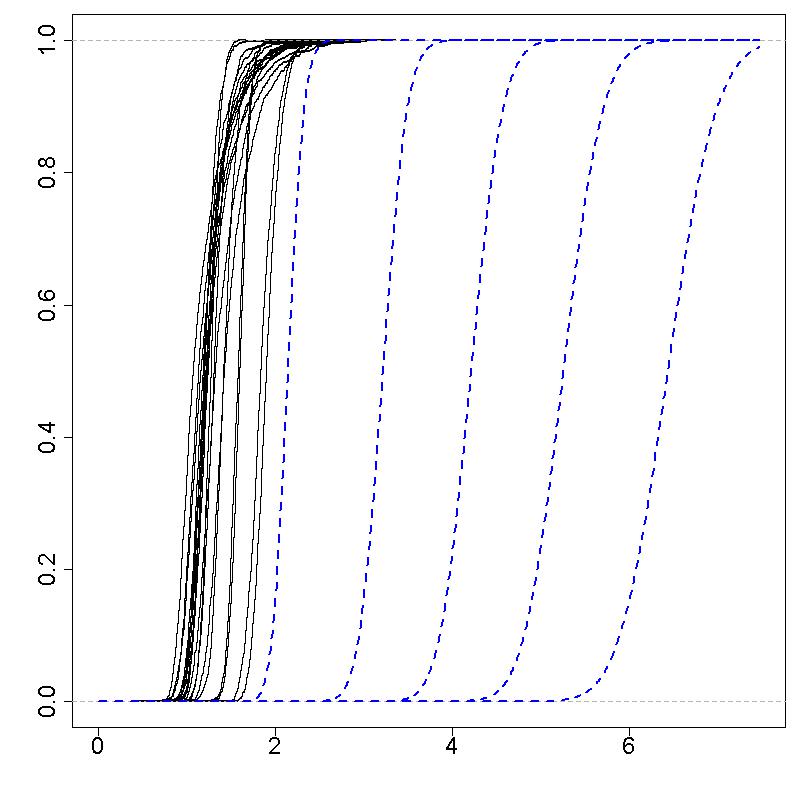}
\caption{Every 5 years}
\end{subfigure}
\begin{subfigure}{0.3\textwidth}
\includegraphics[width=\textwidth]{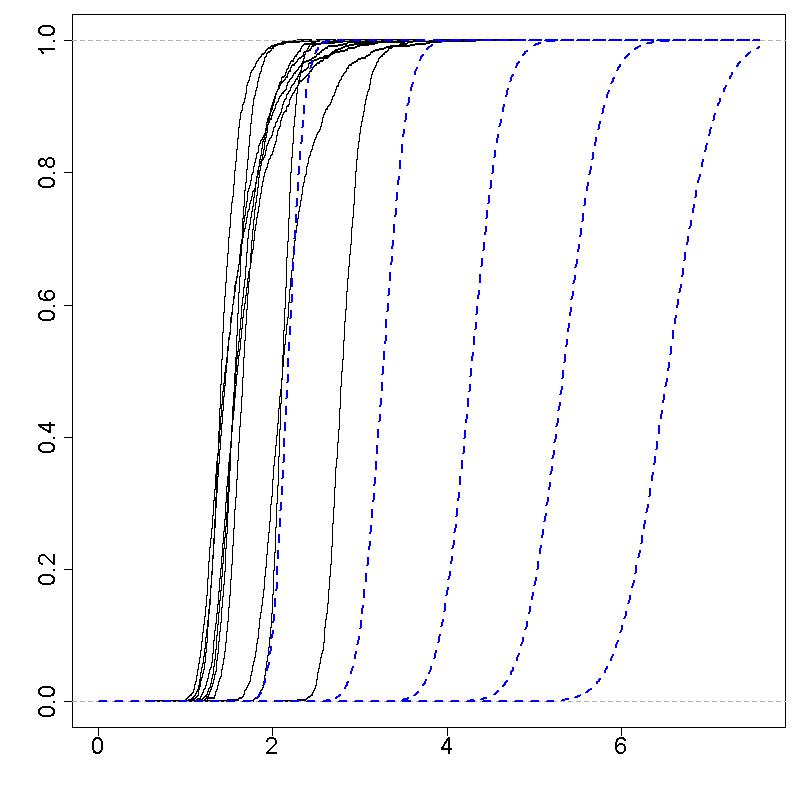}
\caption{Every 10 years}
\end{subfigure}\\

\begin{subfigure}{0.3\textwidth}
\includegraphics[width=\textwidth]{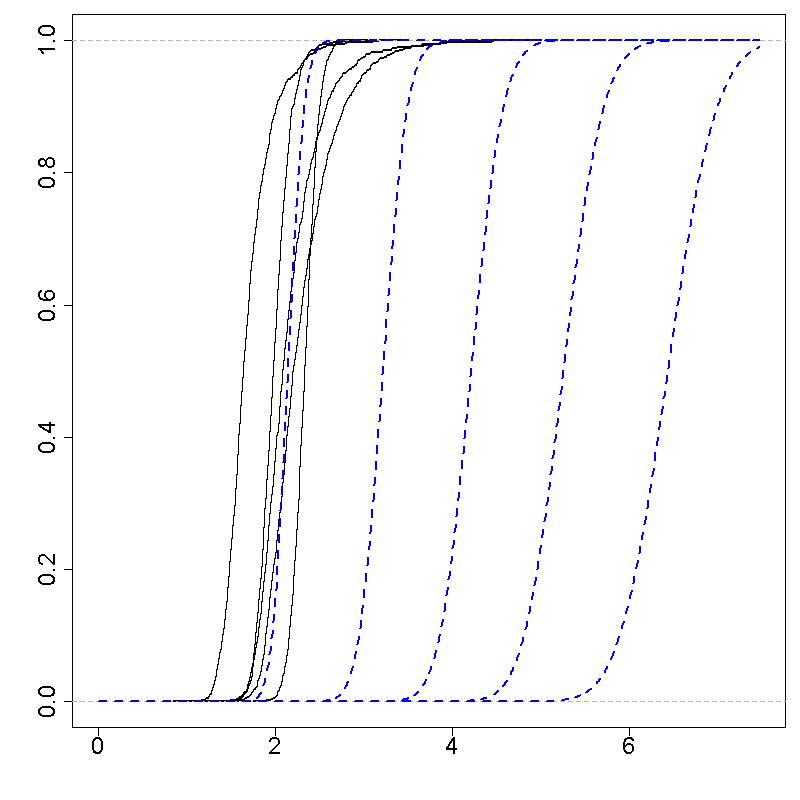}
\caption{Every 20 years}
\end{subfigure}
\begin{subfigure}{0.3\textwidth}
\includegraphics[width=\textwidth]{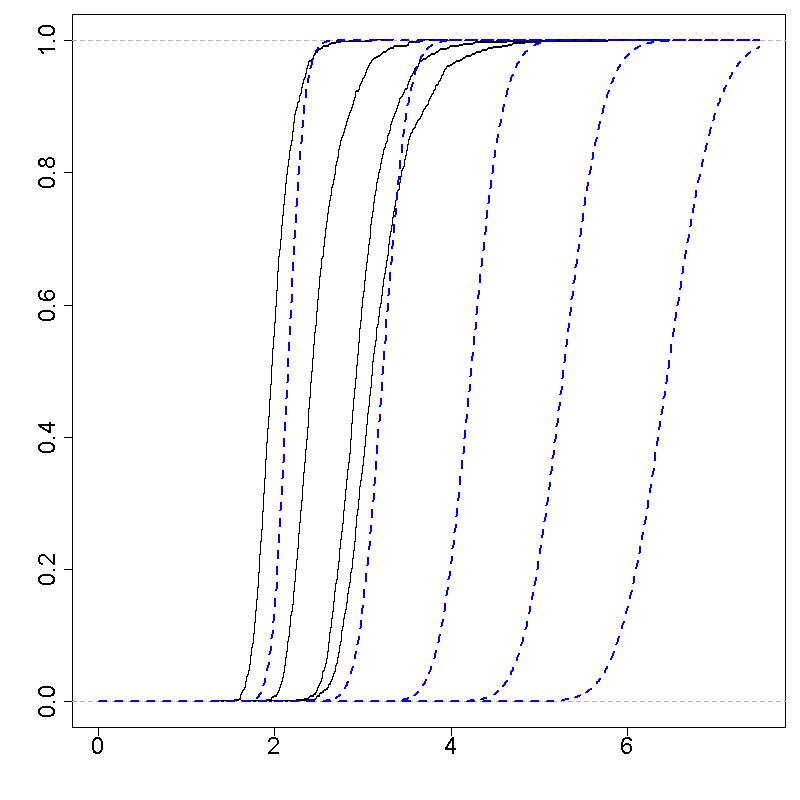}
\caption{Every 25 years}
\end{subfigure}
\caption{World trade data: Plots of $\pbe{t}$ vs. $\epsilon$.  Each curve corresponds to the random effects from a particular time point.  The dotted lines in each figure correspond to the comparative random variable setting $p=0.05,0.1,0.15,0.2,0.25,0.3$.  See Section \ref{evidence} for details.  Coarser time intervals lead to stronger evidence of simultaneous dependence.}
\label{COWSDIncTrend}
\end{figure}

\begin{figure}
\centering
\includegraphics[width=0.45\textwidth]{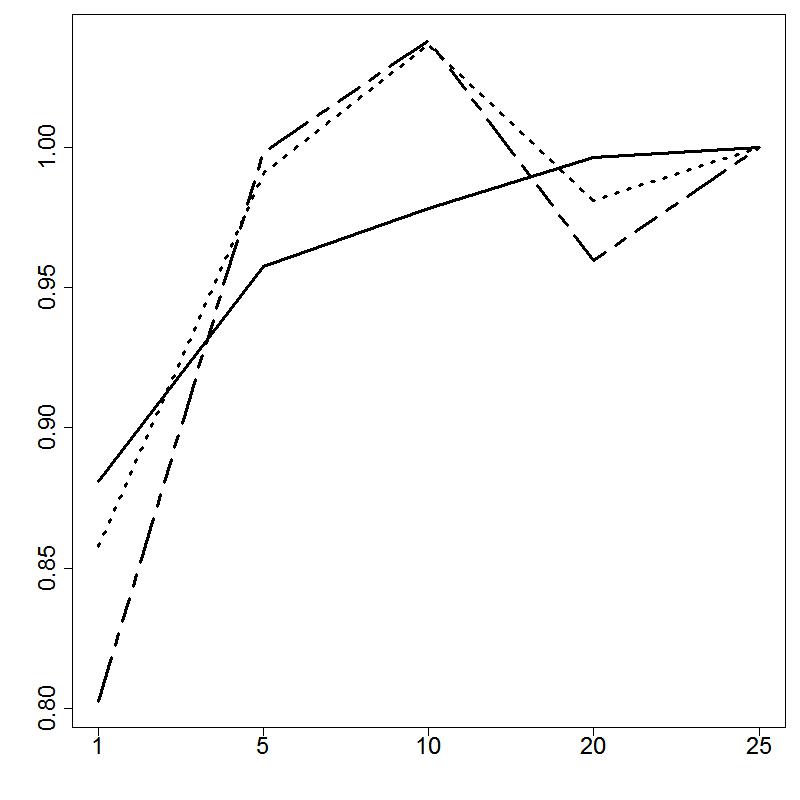}
\caption{TERGM coefficient estimates for reciprocity (solid), transitive triples (dotted), and cyclic triples (dash-dot) (negative coefficients given for the cyclic triples).  Horizontal axis corresponds to the spacing of observations for the data set used.  The increasing trend in the strength of the effect sizes corroborates our finding of increasing simultaneous dependence.}
\label{tergmFig}
\end{figure}

Figure \ref{COWSDIncTrend} gives the evidence of the simultaneous dependence for the five data sets.  We can see that simultaneous dependence increases with the coarseness of the time interval, as shown by the increasing trend for the location of the thin shell of posterior probability mass for the individual effects.  To corroborate this, we also implemented the TERGM model on the five different data sets (collected every 1, 5, 10, 20, and 25 years).  To capture the simultaneous dependencies, we included as ERGM terms the counts of reciprocated ties, transitive triangles, and 3-cycles.  Figure \ref{tergmFig} shows the trends of these parameter estimates for the five data sets, where the values for each parameter have been normalized by the corresponding parameter value from the 25 year interval data.  We see that the strength of the effect sizes increase as the time between observations increases (we actually show the negative of the 3-cycle coefficients for visual clarity), thus corroborating our finding that the simultaneous dependence does in fact increase.

\section{Discussion}
In this paper we have adapted the dynamic logistic network regression model of \cite{almquist2013dynamic} by introducing a framework for capturing not only temporal dependencies through an autoregressive mean structure but also simultaneous dependence through an autoregressive covariance structure.  We demonstrated that ignoring simultaneous dependence leads to negative inferential consequences.  The methods outlined here account for both complex temporal and simultaneous dependencies in a parsimonious way, while keeping within a familiar framework.

Like many other statistical models for network data, scalability is an issue for all but very simple simultaneous dependence structures.  While the VB estimation method proposed for the STAR model is quick for small to medium data sets, the requirement to invert large covariance matrices prohibits this methodology in its current state from being scaled up to extremely large networks.  

We have also described how our work may be placed within the familiar GLMM framework.  While it is beyond the scope of this paper to thoroughly discuss model selection problems involving, e.g., covariance structures or link functions, it is the author's hope that previous and ongoing GLM and GLMM research \citep[e.g.,][]{chen2010probit} can be used to build upon the proposed work in this area.  Further, while we have shown practical operationalizations of the proposed method for binary data in Section \ref{example}, we leave it for future work to describe the specifics of sophisticated covariance structures (i.e., $H_{\cdot,t}$'s that are more complicated than $I_n$) for other data types.

Other future work that would be valuable to the network analysis community would be to provide a thorough comparison of the available methods for discrete temporal network data, such as the proposed approach, TERGM \citep{hanneke2010discrete} and STERGM \citep{krivitsky2014separable}, latent space models for dynamic networks \citep{durante2014nonparametric,sewell2014latent}, and dynamic stochastic blockmodels \citep{xing2010state}.  It would be important to know which method ought to be used in various contexts, and under what circumstances the conclusions from these models might differ.

\appendix
\section{Closed form updates for VB}
\label{VBsolutions}
Before giving the closed form of the $q$'s, let us first provide a little notation that will be used. Let $\Imin = J_n - I_n$, i.e., the matrix of ones with zeros on the diagonal.  Let $tr(A)$ be the trace of some square matrix $A$.  For a matrix $\Sigma$, let $\Sigma_{(i,j)}$ denote the $2\times2$ submatrix obtained from the $i^{th}$ and $j^{th}$ rows and columns.  Let $\cAmin_t$ denote $\vm{A_t^*}$.  Let $trN(\bmu,\Sigma)$ be the truncated normal; we will not add any notation specifying the varying domain as this should be obvious in our context from the data which $A_{ijt}^*$ are restricted to the positive reals and which to the negative reals.  Finally, let $\vec{X}_t$ denote the $n(n-1)\times(p_1+p_2)$ matrix such that
$$
\vec{X}_t = (\vm{X_{1t}},\ldots,\vm{X_{p_1t}},\vm{{\cal G}_{1t}},\ldots,\vm{{\cal G}_{p_2t}}).
$$

\begin{result}
$q_1(\bbeta,\btheta)\eqdist N(\bmu_m,\Sigma_m),$ where
\begin{align*}
\Sigma_m^{-1}&= \diag(1/\sigma^2_{\beta},\ldots,1/\sigma^2_{\beta},1/\sigma^2_{\theta},\ldots,1/\sigma^2_{\theta}) + \sum_{t=1}^T \vec{X}_t'\vec{X},&\\ \nonumber
\bmu_m&=\Sigma_m\left(
\sum_{t=1}^T\vec{X}_t'(M_{A_t} - \vm{(\bmu_{s_1t}+\bmu_{s_2t})\ones'} - \vm{\ones(\bmu_{r_1t}+\bmu_{r_2t})'}-\vm{M_{R_t}})
\right).&
\end{align*}
\label{betaThetaResult}
\end{result}
\vspace{-2pc}
\begin{result}
$q_2(\tau_{s2},\tau_{r2},\Omega)\eqdist IG(a_s,b_s) IG(a_r,b_r) IW(a_{\Omega},B_{\Omega})$ where
\begin{equation*}
\begin{array}{lcl}
a_s=a_{s0} + nT/2 &  
&b_s=b_{s0} + \frac12 \sum_{t=1}^T\left[ tr(\widetilde\Sigma_{srt(s)}H_{st}^{-1}) + \bmu_{s_2t}'H_{st}^{-1}\bmu_{s_2t} \right] \\
a_r=a_{r0} + nT/2 &
& b_r=b_{r0} + \frac12 \sum_{t=1}^T\left[ tr(\widetilde\Sigma_{srt(r)}H_{rt}^{-1}) + \bmu_{r_2t}'H_{rt}^{-1}\bmu_{r_2t} \right] \\
a_{\Omega}=a_{\Omega 0} + nT&
& B_{\Omega}=B_{\Omega 0} + \sum_{t=1}\sum_{i=1}^n \left[ \widetilde\Sigma_{srt(sr)(i,n+i)} + (\bmu_{s_1ti}, \bmu_{r_1ti})'(\bmu_{s_1ti}, \bmu_{r_1ti}) \right],
\end{array}
\end{equation*}
$\widetilde\Sigma_{srt(s)}$ is the first $n$ rows and first $n$ columns of $\widetilde\Sigma_{srt}$, $\widetilde\Sigma_{srt(r)}$ is the second $n$ rows and second $n$ columns of $\widetilde\Sigma_{srt}$, and $\widetilde\Sigma_{srt(sr)}$ is the last $(2n)$ rows and $(2n)$ columns of $\widetilde\Sigma_{srt}$. 

\label{varComponentsResult}
\end{result}

\begin{result}
$q_3(\{\cAmin_t\}_{t=1}^T)\eqdist \prod_{t=1}^T trN(M_{A_t},I)$ where 
\begin{equation*}
M_{A_t}=\vec{X}_t\bmu_m + \vm{(\bmu_{s_1t}+\bmu_{s_2t})\ones'} + \vm{\ones(\bmu_{r_1t}+\bmu_{r_2t})'} +\vm{M_{R_t}}.
\end{equation*}
\label{AstarResult}
\end{result}

\vspace{-2pc}\begin{result} $q_4(\{\bs_{1t},\br_{1t},\bs_{2t},\br_{2t}\}_{t=1}^T)\eqdist\prod_{t=1}^T N\left(
(\bmu_{s_1t}',\bmu_{r_1t}',\bmu_{s_2t}'\bmu_{r_2t}')',
\widetilde\Sigma_{srt}\right)$, where
\begin{align*}
\widetilde\Sigma_{srt}^{-1}&= \left(
\begin{array}{cccc}
1&0&1&0\\0&1&0&1\\ 1&0&1&0\\0&1&0&1
\end{array}\right)\otimes (n-1)I_n + 
\left(
\begin{array}{cccc}
0&1&0&1\\ 1&0&1&0\\0&1&0&1\\ 1&0&1&0
\end{array}\right)\otimes \Imin &\\
&
\hspace{1pc}+  \left(\begin{array}{c|c}
a_{\Omega}B_{\Omega}^{-1}\otimes I_n &0 \\ \hline
0& \begin{array}{cc}
\frac{a_s}{b_s}H_{s1t}^{-1} & 0 \\ 
0 & \frac{a_r}{b_r}H_{r1t}^{-1} 
\end{array} \end{array}\right)&
\end{align*}

\begin{align*}
\left(\begin{array}{c}\bmu_{s_1t}\\ \bmu_{r_1t} \\ \bmu_{s_2t} \\ \bmu_{r_2t}\end{array}\right)&=\widetilde\Sigma_{srt}\left(\begin{array}{c}
     \left(  \revvm{M_{A_t}-\vec{X}_t\bmu_m}-M_{R_t}\right)\ones 
\\   \left(  \revvm{M_{A_t}-\vec{X}_t\bmu_m}'-M_{R_t}\right)\ones 
\\   \left(  \revvm{M_{A_t}-\vec{X}_t\bmu_m}-M_{R_t}\right)\ones 
\\   \left(  \revvm{M_{A_t}-\vec{X}_t\bmu_m}'-M_{R_t}\right)\ones 
\end{array}\right)&
\end{align*}
and $\revvm{\cdot}$ is the matrix (with zero diagonal elements) constructed by reversing the $\vm{\cdot}$ operator.

\label{REResults}
\end{result}
\deriv{

We first provide some preliminary results:
\begin{enumerate}
\item For some $n\times1$ vectors ${\bf a}_1$ and ${\bf a}_2$, $tr(D_{a_1}\Imin \Imin  D_{a_2} )=(n-1){\bf a}_1'{\bf a}_2$, where $D_a$ denotes a diagonal matrix whose entries are ${\bf a}$.
\item For some $n\times n$ matrix $A$, $tr(\Imin D_a (A\circ \Imin))={\bf a}'(A\circ \Imin)\ones$.
\item $tr(\Imin D_{a_1} \Imin D_{a_2})= {\bf a}_1'\Imin {\bf a}_2$. 
\end{enumerate}
Also note that since $\vm{A}'\vm{A}=\mvec(A\circ \Imin)'\mvec(A\circ \Imin)=tr((A\circ \Imin)'(A\circ \Imin))$, we may consider the conditional probability of $\cA_t|\bs_{1t},\br_{1t},\bs_{2t}\br_{2t},\cdot$ as proportional (with respect to the sender and receiver effects) to the matrix normal distribution kernel of $A_t^*\circ \Imin$.

Letting $\tilde{A}_t=(A_t^*-\langle\bbeta,\cX_t\rangle + \langle\btheta,{\cal G}_t\rangle)\circ \Imin$, we have, dropping the subscript $t$,
\begin{align*}
&\log(\pi(A^*|\bs_{1},\br_{1},\bs_{2},\br_{2},\cdot)) &\\
&=\const-\frac12tr\left[
(\tilde{A}-D_{s1}\Imin -D_{s2}\Imin-\Imin D_{r1}-\Imin D_{r2})'(\tilde{A}-D_{s1}\Imin -D_{s2}\Imin-\Imin D_{r1}-\Imin D_{r2})
\right]&\\
&=\const-\frac12tr\left[
\Imin D_{s1}D_{s1} \Imin - 2\Imin D_{s1}\tilde{A}+2\Imin D_{s1}D_{s2}\Imin +2\Imin D_{s1}\Imin D_{r1} + 2\Imin D_{s1}\Imin D_{r2}\right. &\\
& - 2\Imin D_{s2}\tilde{A} +\Imin D_{s2}D_{s2}\Imin + 2\Imin D_{s2}\Imin D_{r1} + 2\Imin D_{s2}\Imin D_{r2} +D_{r1}\Imin \Imin D_{r1} + 2D_{r1}\Imin \Imin D_{r2} &\\
&\left.+ D_{r2}\Imin \Imin D_{r2} -2D_{r1}\Imin \tilde{A} -2D_{r2}\Imin\tilde{A}
\right]&\\
&=\const-\frac12\left[
(n-1)\bs_1'\bs_1 -2\bs_1\tilde{A}\ones + 2(n-1)\bs_1'\bs_2+2\bs_1'\Imin\br_1 + 2\bs_1'\Imin\br_2 -2\bs_2'\tilde{A}\ones + (n-1)\bs_2'\bs_2 \right.&\\
&\left. + 2\bs_2'\Imin\br_1 + 2\bs_2'\Imin\br_2 + (n-1)\br_1'\br_1+2(n-1)\br_1'\br_2+(n-1)\br_2'\br_2 -2\br_1'\tilde{A}'\ones - 2\br_2'\tilde{A}'\ones
\right].&\\
\end{align*}
Combining the expected value of this under $q$ with $\E_q\left(\log(\pi(\bs_{1t},\br_{1t},\bs_{2t},\br_{2t}|\tau_{s2},\tau_{r2},\Omega,A_{t-1}))\right)$ yields Result \ref{REResults}.
}

\begin{result}
$q_5(\{R_t\}_{t=1}^T)\eqdist\prod_t\prod_{i<j}N(M_{R_t}[i,j],\widetilde\sigma_R^2)$ where
\begin{align*}
M_{R_t}[i,j]&=\widetilde\sigma^2_R(\widetilde A_{ijt}+\widetilde A_{jit}),&\\
\widetilde\sigma_R^2&=\frac{b_R/a_R}{1+2b_R/a_R},&\\
\widetilde A_{ijt} &=  \revvm{M_{A_t}-\vec{X}_t\bmu_m}[i,j] -\bmu_{s_1t}[i]-\bmu_{s_2t}[i]-\bmu_{r_1t}[j]-\bmu_{r_2t}[j].&
\end{align*}
For the purposes of computing the parameters for the other $q$'s, assume for $i<j$ that $M_{R_t}[j,i]=M_{R_t}[i,j]$.
\end{result}

\begin{result}
$q_6(\sigma^2_R)\eqdist IG\left( a_R,b_R \right)$ where
\begin{align*}
a_R&=a_{R0}+\frac{Tn(n-1)}{4}&\\
b_R&=b_{R0}+\frac12\sum_t\sum_{i<j}\left( \widetilde\sigma^2_R + M_{R_t}[i,j]^2 \right)
\end{align*}
\end{result}

\begin{result}
\label{resultREUndir}
For the undirected case, $q_4(\{\bs_t\}_{t=1}^T)=\prod_{t=1}^T N(\bmu_{st}',\widetilde\Sigma_{st})$, where
\begin{align*}
\bmu_{st}&=\widetilde\Sigma_{st}\E(A_t^*\circ\Imin)\ones&\\
\widetilde\Sigma_{st}^{-1}&= (n-1)I_n + \Imin + \frac{a_s}{b_s}H_{st}^{-1}&
\end{align*}
\end{result}
\deriv{
Define $\Itri$ as the square matrix with ones on the upper triangle and zero everywhere else (the diagonal is also zero).  As before, it is helpful to provide some preliminary results:
\begin{enumerate}
\item For some $n\times1$ vector {\bf a}, $tr(D_{{\bf a}}(\Itri\Itri' + \Itri'\Itri){\bf a})=(n-1){\bf a}'{\bf a}$.
\item For some $n\times n$ matrix $A$, $$tr(D_{{\bf a}}(\tilde{A}'\Itri + \tilde{A}\Itri'))=tr(D_{{\bf a}}\Imin A) = {\bf a}'(A\circ\Imin)\ones.$$
\item $2\cdot tr(D_{{\bf a}}\Itri'D_{\bf a}\Itri)={\bf a}'\Imin{\bf a}$.
\end{enumerate}
To show this last, note that the $i^{th}$ diagonal of $D_{{\bf a}}\Itri'D_{\bf a}\Itri=\sum_{j=1}^{i-1}{\bf a}_i{\bf a}_j$, and hence the trace equals $\sum_{i=1}^n\sum_{j=1}^{i-1}{\bf a}_i{\bf a}_j={\bf a}'\Itri'{\bf a}={\bf a}'\Itri{\bf a}$.  This then implies that $2\cdot tr(D_{{\bf a}}\Itri'D_{\bf a}\Itri)={\bf a}'\Itri'{\bf a}+{\bf a}'\Itri{\bf a}={\bf a}'\Imin{\bf a}$.

Let $\tilde{A_t}=(A_t^*-\langle\bbeta,\cX_t\rangle + \langle\btheta,{\cal G}_t\rangle)\circ\Itri$.  Then we have, dropping the subscript $t$,
\begin{align*}
&\log(\pi(A_t^*|\bs))&\\
&=\const-\frac12tr\left[
(\tilde{A}-D_s\Itri - \Itri D_s)'(\tilde{A}-D_s\Itri - \Itri D_s)
\right]&\\
&=\const-\frac12tr\left[
D_s(\Itri\Itri' + \Itri'\Itri)D_s + 2D_s\Itri' D_s\Itri -2D_s(\tilde{A}'\Itri + \tilde{A}\Itri')
\right]&\\
&\const-\frac12\left[
\bs'\Big((n-1)I+\Imin+\frac{1}{\tau_s}H_s^{-1}\Big)\bs - 2\bs'(A^*\circ\Imin)\ones
\right].
\end{align*}
Combining the expected value of this under $q$ with $\E_q(\log(\pi(\bs_t|\tau_s,A_{t-1})))$ yields Result \ref{resultREUndir}.}

\section{Proofs}
\subsection{Proposition of Section \ref{estimationErrors}}
\label{proofOfProposition}
\begin{proof}
Letting $m_{ijt}=\langle \bbeta,\cX_t\rangle[i,j] + \langle \btheta,{\cal G}_t\rangle[i,j]$ and $V:=Var(s_{it}+r_{jt})$, we have
\begin{align*}
\Prob(A_{ijt}=1|\bbeta,\btheta)&=\E\Big(\E\big(A_{ijt}\big|s_{it}+r_{jt},\bbeta,\btheta\big)\big|\bbeta,\btheta\Big)&\\
&=\E\left( \boldsymbol\Phi\left(\frac{s_{it}+r_{jt}+m_{ijt}}{\sqrt{Var(E_{ijt})}}\right)\Big|\bbeta,\btheta\right)&\\
&=\int_{-\infty}^\infty\int_{-\infty}^{\frac{s_{it}+r_{jt}+m_{ijt}}{\sqrt{Var(E_{ijt})}}}\frac{1}{\sqrt{2\pi}}e^{-\frac{Z^2}{2}}\frac{1}{\sqrt{2\pi V}}e^{-\frac{(s_{it}+r_{jt})^2}{2V}}dZd(s_{it}+r_{jt})&\\
&=\Prob(Z\sqrt{Var(E_{ijt})} - (s_{it}+r_{jt})<m_{ijt}).&
\end{align*}
Since $Z\sqrt{Var(E_{ijt})} - (s_{it}+r_{jt})\sim N(0,Var(E_{ijt})+V)$, our result holds.
\end{proof}

\subsection{Theorem of Section \ref{STARModel}}
\label{proofOfTheorem}
\begin{proof}

It is obvious that the mean of each $A_{ijt}^*$ are equivalent for (I), (II), and (III), and that the covariance between any $A_{ijt}^*$ and $A_{k\ell t}^*$ as given by (III) satisfies (\ref{rbacs1}).  

It is straightforward to check that $\sigma^2_RM_R+(\sigma^2_\epsilon+\sigma^2_R)I_{n^2}$ satisfies the final two terms in (\ref{rbacs1}), and that this is the covariance matrix of $\mvec(E_t)$.  Note that for any two $n$-dimensional vectors ${\bf a}$ and ${\bf b}$, we have that
\begin{enumerate}[(i)]
\itemsep0em
\item  $\mbox{vec}({\bf a}{\bf b}') = {\bf b}\otimes{\bf a}$,
\item $\cov(\ones\otimes{\bf a})= J_n\otimes \cov({\bf a})$,
\item $\cov({\bf a}\otimes\ones)= \cov({\bf a})\otimes J_n$, and
\item $\cov(\ones\otimes{\bf a},{\bf b}\otimes\ones)=\ones\otimes \cov({\bf a},{\bf b})\otimes \ones'$,
\end{enumerate}
where $J_n$ is the $n\times n$ matrix of 1's.  We may then write the covariance of the $A_{ijt}^*$'s as given in (III) as 
\begin{align*}\nonumber
\cov(\cA_t)&=\cov(\mvec(\bs_t\ones')+\mvec(\ones\br_t')+\mvec(E_t))&\\ \nonumber
&=\cov(\ones\otimes\bs_t +\br_t\otimes\ones+\mvec(E_t))&\\
&= J_n\otimes\Sigma_{st} + \Sigma_{rt}\otimes J_n + \ones\otimes\Sigma_{srt}\otimes\ones' + \ones'\otimes \Sigma_{srt}'\otimes\ones + \sigma^2_RM_R+(\sigma^2_\epsilon+\sigma^2_R)I_{n^2}.&
\end{align*}
Hence (I), (II), and (III) have the same covariance structure.

Finally, we have from (III)
\begin{align*}
\cA_t&=\mvec(\langle\bbeta,\cX_t\rangle + \langle\btheta,{\cal G}_t\rangle) +\ones\otimes\bs_t +\br_t\otimes\ones + \mvec(E_t)&\\
&=\mvec(\langle\bbeta,\cX_t\rangle + \langle\btheta,{\cal G}_t\rangle )+\big(\ones\otimes I_n\big)\bs_t + \big(I_n\otimes\ones\big)\br_t + \mvec(E_t)&\\
&\eqdist \mvec(\langle\bbeta,\cX_t\rangle + \langle\btheta,{\cal G}_t\rangle) +\Big(  
\big(\ones\otimes I_n, I_n\otimes\ones\big)\Sigma_t^{\frac12},\big(\sigma^2_RM_R+(\sigma^2_\epsilon+\sigma^2_R)I_{n^2}\big)^{\frac12}
\Big) {\bf z}
\end{align*}
where ${\bf z}$ is a $(2n+n^2)\times1$ vector of independent standard normal random variables, and
$$
\Sigma_t := \left(\begin{array}{cc} \Sigma_{st} &\Sigma_{srt}\\ \Sigma_{srt}' & \Sigma_{rt} \end{array}\right),
$$
Since $\mvec(\cA_t)$ is an affine transformation of ${\bf z}$, we have that the $A_{ijt}^*$'s are jointly normal, 
indicating that (I), (II), and (III) are equivalent.
\end{proof}

\bibliographystyle{nws}
\bibliography{StarBib1}

\end{document}